\documentclass[twocolumn]{aastex63}

\usepackage{bm}
\usepackage[T1]{fontenc}

\usepackage{grffile} 
\newcommand{\amon}{a_\mathrm{mon}}
\newcommand{\amin}{a_\mathrm{min}}
\newcommand{\amax}{a_\mathrm{max}}

\received{January 11, 2023}
\revised{--}
\accepted{--}
\submitjournal{ApJ}

\shorttitle{Phase functions of planet-forming disks}
\shortauthors{Ginski et al.}
\graphicspath{{./}{figures/}}

\begin{document}
\title{Observed polarized scattered light phase functions of planet-forming disks}

\correspondingauthor{Christian Ginski}
\email{ginski@strw.leidenuniv.nl}

\author[0000-0002-4438-1971]{Christian Ginski}
\affiliation{Leiden Observatory, Leiden University, PO Box 9513, 2300 RA Leiden, The Netherlands}
\affiliation{Anton Pannekoek Institute for Astronomy, University of Amsterdam, Science Park 904, 1098XH Amsterdam, The Netherlands}

\author[0000-0003-1451-6836]{Ryo Tazaki}
\affiliation{Anton Pannekoek Institute for Astronomy, University of Amsterdam, Science Park 904, 1098XH Amsterdam, The Netherlands}
\affil{Institute of Planetology and Astrophysics, Université Grenoble Alpes, 38000 Grenoble, France}
\affil{Astronomical Institute, Graduate School of Science,
Tohoku University, 6-3 Aramaki, Aoba-ku, Sendai 980-8578, Japan}

\author[0000-0002-3393-2459]{Carsten Dominik}
\affiliation{Anton Pannekoek Institute for Astronomy, University of Amsterdam, Science Park 904, 1098XH Amsterdam, The Netherlands}

\author[0000-0002-5823-3072]{Tomas Stolker}
\affiliation{Leiden Observatory, Leiden University, PO Box 9513, 2300 RA Leiden, The Netherlands}



\begin{abstract}

Dust particles are the building blocks from which planetary bodies are
made.  A major goal of the studies of planet-forming disks is to
constrain the properties of dust particles and
aggregates in order to trace their origin, structure, and the associated growth
and mixing processes in the disk. Observations of scattering and/or emission of dust in a
location of the disk often lead to degenerate information
about the kind of particles, such as size, porosity, or fractal
dimension of aggregates. Progress can be made by deriving the full
(polarizing) scattering phase function of such particles at multiple
wavelengths. This has
now become possible by careful extraction from scattered light
images. Such an extraction requires knowledge about the shape of the
scattering surface in the disk and we discuss how to obtain such
knowledge as well as the associated uncertainties.
We use a sample of disk images from observations with
VLT/SPHERE to, for the first time, extract the phase functions of
a whole sample of disks with broad phase angle coverage. We find that
polarized phase functions come in two categories. Comparing the extracted
functions with theoretical predictions from rigorous T-Matrix
computations of aggregates, we show that one category can be linked
back to fractal, porous aggregates, while the other is consistent
with more compact, less porous aggregates. We speculate that the more
compact particles become visible in disks where embedded planets
trigger enhanced vertical mixing.

\end{abstract}



\section{Introduction}

Gas- and dust-rich circumstellar disks are the sites of ongoing planet formation. However, the early stage of planet formation, such as the formation of planetesimals, remains a matter of discussion, as it comes with a number of barriers inhibiting grain growth and forming planetesimals (e.g., \citealt{Brauer2008, Zsom2010, Birnstiel2012}). The sizes and structures of growing dust aggregates are of crucial relevance to the barriers because these quantities influence the sticking and aerodynamic properties (\citealt{Ormel2007, Zsom2010, Okuzumi2012, Kataoka2013, Krijt2016, Lorek2018, Garcia2020, Kobayashi2021, Estrada2022}) Thus by determining these properties by disk observations, one can conclude the early planet formation and transport processes including vertical mixing.

The past decade was revolutionary for resolved observations of young planet-forming disks in the near-infrared. Driven by advances in instrumentation several large surveys have been conducted or are still ongoing such as SEEDS (\citealt{Tamura2016}), DARTTS-S (\citealt{Avenhaus2018, Garufi2020}), Gemini-LIGHTS (\citealt{Rich2022AJ}) and SPHERE-DESTINYS (\citealt{Ginski2021}). A summary of the field was recently presented by \cite{Benisty2022}. Due to these ongoing observational programs more than 150 systems have now been observed in (polarized) near-infrared scattered light. 
A large diversity of sub-structures has been discovered, such as rings, gaps and spirals, which are typically associated with ongoing planet formation (see e.g. \citealt{Benisty2022}). The analysis of the scattered light data often focused on either the basic disk geometry, tracing illumination and shadowing (e.g. \citealt{Garufi2022}) as a function of disk aspect ratio and disk symmetry, or on the disk morphology (e.g. \citealt{Garufi2018}). 

However, the appearance of these objects in (polarized) scattered light is also strongly dependent on the properties of the dust grains or aggregates in the upper disk atmosphere (see e.g. \citealt{Min2005}). In particular the scattering angle dependent amount of flux that we receive from the different regions of the disk, the so-called (polarized) scattering phase function, encodes dust grain and aggregate properties (\citealt{Tazaki2019}). While the extraction of polarized scattered light phase functions has been done for geometrically flat debris disks (e.g. \citealt{Milli2017,Milli2019,Olofsson2020,Engler2022}), it is less common for young gas rich disks, due to their more complex geometry. So far this has only been done for the disks around the Herbig stars HD\,97048 (\citealt{Ginski2016}) and HD\,100546 (\citealt{Quanz2011, Stolker2016}) in polarized light and the T Tauri star multiple system GG\,Tau (\citealt{McCabe2002}) in total intensity. 

In this study we gathered a sample of 10 disks for which the surface \textit{height} profile has been determined in the literature from near-infrared scattered light observations. 
The sample is comprised of Herbig and T Tauri stars with the earliest spectral type being the B9.5 star HD\,34282 and the latest spectral type the M0 star IM\,Lup. The systems cover a range of ages from 1.1\,Myr (IM\,Lup, \citealt{Avenhaus2018}) to 12.7\,Myr (MY\,Lup, \citealt{Avenhaus2018}). All systems were also already observed at (sub)mm-wavelengths which allowed to estimate dust masses based on their continuum flux. Our sample spans roughly an order of magnitude between the lowest mass disk around PDS\,70 (13\,M$_\oplus$) and the highest mass disk in the RXJ\,1615.3-3255 system with (140\,M$_\oplus$).  
A summary of all included systems with the appropriate references is given in table~\ref{tab: system-summary}.
Due to the ring-shaped sub-structure in most of these disks, planet formation is thought to be ongoing. In particular our study includes the PDS\,70 system, in which two young planets have been detected inside the disk cavity (\citealt{Keppler2018, Haffert2019}), and the HD\,163296 system for which the presence of two wide separation planets has been inferred from ALMA gas kinematic observations (\citealt{Pinte2018, Teague2018}). 
In the following section we briefly describe the observational data. In section~\ref{sec: extraction} we discuss how the phase functions were extracted, while we discuss their interpretation in light of dust aggregate models in section~\ref{sec: discussion}. We summarize our results in section~\ref{sec: summary}.

\begin{table}[h!]
\centering
\begin{tabular}{lccc}
\hline
System & SpType & M$_{dust}$ (M$_{\oplus}$) & Ref. \\
\hline
RXJ\,1615.3-3255 & K5 & 140& (1),(12) \\
HD\,163296 & A1 & 75 & (2),(3) \\
IM\,Lup & M0 & 54 & (1),(3) \\
LkCa\,15 & K5.5 & 33 & (4),(3) \\
PDS\,66 & K1.5 & 15 & (5),(3) \\
PDS\,70 & K7 & 12 & (1),(3) \\
2MASSJ18521730-3700119 & K4 & 13 & (6),(3) \\
V\,4046\,Sgr & K4 & 48 & (7),(10) \\
HD\,34282 & B9.5 & 87 & (8),(11) \\
MY Lup & K0 & 53 & (1),(9) \\

\hline
\hline
\end{tabular}
\label{tab: system-summary}
\caption{(1) \cite{Luhman2022} (2) \cite{Sartori2003} (3) \cite{Marel2021} (4) \cite{Krolikowski2021} (5) \cite{Pecaut2016}
(6) \cite{Herczeg2014} (7) \cite{Pecaut2013} (8) \cite{Kharchenko2001}
(9) \cite{Mulders2017} (10) \cite{Martinez-Brunner2022} (11) \cite{Stapper2022} (12) \cite{Marel2019}}
\end{table}

\section{Observations and data reduction}

All datasets used in our study have been obtained with VLT/SPHERE (\citealt{Beuzit2019}) and its near infrared camera IRDIS (\citealt{Dohlen2008}). The instrument was operated in dual-beam polarization imaging mode (DPI, \citealt{deBoer2020,vanHolstein2020}), in either J or H broad band filters, to obtain (linear) polarized scattered light images of the circumstellar disks in each system. An overview of the observation setup and conditions is given in table~\ref{tab: observations} in appendix~\ref{app: observations}. In all cases the innermost 92.5\,mas around the stellar position where covered by the standard \emph{YJH\_S} apodized Lyot coronagraph (\citealt{Carbillet2011}) and are therefore inaccessible for the analysis. 
All data sets that are included in our study have been previously discussed in the literature. We give the relevant references in table~\ref{tab: surface-profiles}. 
The data reduction has in all cases been carried out with the public IRDAP (IRDIS Data reduction for Accurate Polarimetry, \citealt{vanHolstein2020}) pipeline with default settings\footnote{https://irdap.readthedocs.io}. This includes a full model based determination and removal of instrumental polarization, as well as the measurement and subsequent subtraction of astrophysical stellar polarization. 

\section{Phase function extraction}
\label{sec: extraction}

\subsection{Surface height profiles}

\begin{figure}
\center
\includegraphics[width=0.49\textwidth]{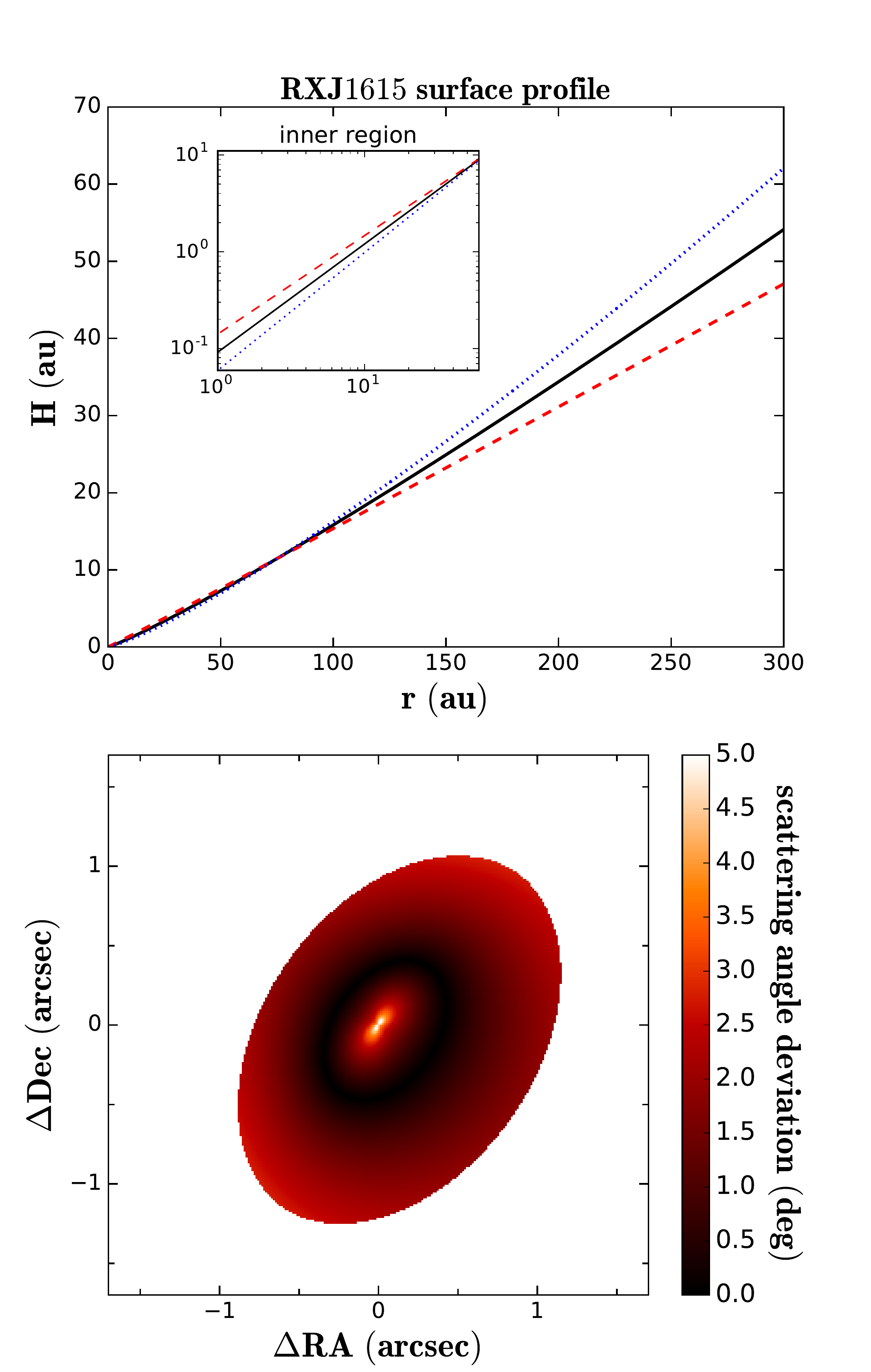} 
\caption{\textit{Top:} Exemplary $\tau$=1 surface profile for the RXJ\,1615 system, as used in our phase function extraction. The solid black curve indicates the nominal surface profile, while the blue-dotted and red-dashed curves indicate the flared and flat extremes, respectively. We consider theses as the boundaries for the region of uncertainty. The inset shows the region inside 50\,au on a log scale for clarity. \textit{Bottom:} Maximum deviation of the scattering angles between the flared and flat disk profiles for each position within the image of the RXJ\,1615 system. We find the largest deviation of up to 5$^\circ$ close to the star and significantly smaller deviations further out. Note that the dark region with the smallest deviations was used for normalization of the surface power-law profile.
} 
\label{fig:surface profile rxj1615}
\end{figure}

The key challenge in extracting the scattered light phase function of young gas-rich disks is the uncertainty of the vertical structure. Here we are particularly interested at which height above the disk mid-plane the optical depth $\tau$ becomes equal to 1, i.e. the surface layer from which the majority of the disk scattered light originates. For ease of use within the further discussion we will refer to this as the surface height of the disk within this study.\footnote{We note that this is not identical to the pressure scale height of the disk, which is typically a factor 3-4 smaller than the scattered light surface height (\citealt{Chiang2001}).}\\
While the surface height may generally be inferred from detailed radiative transfer modeling, there exists a sub-class of disks for which it can be directly determined from the data itself. As shown by \cite{deBoer2016} and \cite{Ginski2016} the disk surface height can be computed for disks with radial sub-structures such as multiple rings. This is done by measuring the inclination of the ring and its offset along the minor axis from the central star position in the image. This directly gives the surface height at the ring location. If there are multiple rings then the radial dependence of the disk surface height can be directly traced.
Both of the mentioned studies found that the surface height profile for the two studied target systems (RXJ\,1615 and HD97048) can be described reasonably well with a single power law profile of the form:

\begin{equation}
H(r) = H_{\mathrm{ref}}/(r_{\mathrm{ref}}/1\mathrm{au})^\alpha * (r/1\mathrm{au})^\alpha ,
\end{equation}
\\
wherein $H(r)$ is the radial dependent surface height, $H_{\mathrm{ref}}$ is a reference height at reference separation $r_{\mathrm{ref}}$ (all in au) and $\alpha$ is the flaring exponent. 
\cite{Avenhaus2018} found that the surface height profile for five disks in their study could all be described by the same power-law profile with a flaring exponent of $\alpha=1.22\pm0.03$. This indicates that for single-ringed disks, i.e. when only a surface height at a single radial separation is known, we may still infer a reasonable guess of the radial dependent surface height profile by using the height measurement as reference height in the power-law and by assuming a standard flaring exponent of $\alpha=1.22$. \\
For our sample we draw the surface profile parameters from the literature. For the RXJ\,1615, IM\,Lup and V\,4046\,Sgr multi-ring systems we use the specific power-law profiles fitted by \cite{Avenhaus2018}. For the HD\,34282 multi-ring system we likewise use the power-law profile given by \cite{deBoer2020}. For the remaining single-ringed disks we use the literature values for the surface height of the rings as reference height and the standard flaring exponent of $\alpha=1.22$ by \cite{Avenhaus2018}. For the RXJ1852 and PDS\,70 systems a different flaring exponent was infered from radiative transfer modelling by \cite{Villenave2019} and \cite{Keppler2018}, respectively, and we make use of their fitted values. \\
In order to capture the uncertainty of the extracted phase functions due to the uncertainty of the surface height we always consider three scenarios for each system: (1) the nominal surface height profile, (2) a strongly "flared" profile and a (3) "flat" profile. The flared and flat profile consider the uncertainty of the reference height, the reference separation and the flaring exponent. For the "flared" profile we use the upper bound on the reference height, the lower bound on the reference separation and the upper bound on the flaring exponent, while we switch lower and upper bounds for each parameter for the "flat" profile. We summarize all profile parameters for each system in table~\ref{tab: surface-profiles} and illustrate all three profiles (nominal, flared and flat) for the RXJ\,1615 system in figure~\ref{fig:surface profile rxj1615}, top panel. In the bottom panel of figure~\ref{fig:surface profile rxj1615} we show the corresponding deviations in scattering angle between the "flared" and "flat" surface profile extremes. While we see deviations of up to $\sim$5$^\circ$ in the inner disk region, these are smaller in the outer disk. We note that the intermediate region between inner disk and outer edge, which shows deviations of less than 1$^\circ$ is close to the reference separation for which the reference height was directly measured by \cite{Avenhaus2018}. Thus the deviation in this region is dominated by the (small) uncertainties of these quantities.
We show similar figures for the maximum deviation of the scattering angle for our complete sample in figure~\ref{fig:angle gallery} in the appendix. Unsurprisingly the largest deviations are found close to the outer disk edge for the systems with large uncertainties in their flaring exponent, i.e. IM\,Lup and HD\,34282 in particular. In general we are avoiding these outer disk regions for phase function extraction and typically consider regions for which the uncertainty in scattering angle is less than $\sim$10$^\circ$, with the main exception being the two aforementioned systems. However, as we explain in the following section we do for all systems incorporate the resulting deviations of the extracted phase functions between all three surface height profiles in their uncertainties.

\begin{table*}[h!]
\centering
\begin{tabular}{lcccccccc}
\hline
System                  &           &   d       & $i$	        & PA	        & $h/r_{\mathrm{ref}}^\alpha$ 	& $r_{\mathrm{ref}}$	& $\alpha$  & Ref.\\
                        &           & $(\mathrm{pc})$    & $(^{\circ})$  & $(^{\circ})$  &           & $(\mathrm{au})$    &           &       \\
\hline
 RXJ\,1615.3-3255       & nominal   &  155.6    &   47.2        & 145.0         & 0.091     & 161.8     & 1.12      & (3)   \\
                        & min       &           &               &               & 0.140     & 162.3     & 1.02      &       \\
                        & max       &           &               &               & 0.059     & 161.3     & 1.22      &       \\
 HD\,163296             & nominal   & 101.0     &  46.0         & 134.8         & 0.086     & 63.6      & 1.22       & (1),(2),(3)       \\
                        & min       &           &               &               & 0.081     & 68.7      & 1.19      &       \\
                        & max       &           &               &               & 0.090     & 58.6      & 1.25      &       \\
 IM\,Lup                & nominal   & 155.8     &  55.0         & 325.0         & 0.046     & 149.6     & 1.27      & (3)   \\
                        & min       &           &               &               & 0.095     & 154.3     & 1.07      &       \\
                        & max       &           &               &               & 0.022     & 144.9     & 1.47      &       \\
 LkCa\,15               & nominal   & 157.2     & 50            & 60            & 0.074     & 57.7      & 1.22      & (3),(4),(5)    \\
                        & min       &           &               &               & 0.059     & 60.1      & 1.19      &       \\
                        & max       &           &               &               & 0.089     & 53.7      & 1.25      &       \\
 PDS\,66                & nominal   &   97.9    &   30.3        & 189.2         & 0.052     & 84.2      & 1.22      & (3)    \\
                        & min       &           &               &               & 0.055     & 84.7      & 1.19      &       \\
                        & max       &           &               &               & 0.050     & 83.5      & 1.25      &       \\
 PDS\,70                & nominal   &   112.4   &   49.7        &  158.6        & 0.041     & 99.1      & 1.25      & (6),(7)      \\
                        & min       &           &               &               & 0.044     & 105.1     & 1.22      &       \\
                        & max       &           &               &               & 0.039     & 93.2      & 1.28      &       \\
 2MASSJ18521730-3700119 & nominal   &  147.1    & 30            & 124           & 0.046     & 43.4      & 1.10      & (8)   \\
                        & min       &           &               &               & 0.065     & 45.2      & 1.00      &       \\
                        & max       &           &               &               & 0.034     & 41.6      & 1.20      &       \\
V\,4046\,Sgr            & nominal   & 71.5      &  32.2         & 74.7          & 0.017     & 26.7      & 1.61      & (3)   \\
                        & min       &           &               &               & 0.027     & 26.8      & 1.47      &       \\
                        & max       &           &               &               & 0.012     & 26.6      & 1.74      &       \\   
HD\,34282               & nominal   & 308.6     &  57.0         & 118.0         & 0.064     & 86.4      & 1.35      & (9)   \\
                        & min       &           &               &               & 0.071     & 89.5      & 1.27      &       \\
                        & max       &           &               &               & 0.057     & 83.3      & 1.43      &       \\  
MY\,Lup                 & nominal   & 157.2     &  77.0         & 239.0         & 0.073     & 121.0     & 1.22      & (3)   \\
                        & min       &           &               &               & 0.073     & 125.8     & 1.19      &       \\
                        & max       &           &               &               & 0.072     & 116.3     & 1.25      &       \\  
\hline
\hline
\end{tabular}
\caption{Values for the surface height geometry of the studied disks. We give the nominal values used as well as the values for the minimal and maximal surface height that we considered. References are: (1) \cite{Muro-Arena2018} (2) \cite{Isella2016} (3) \cite{Avenhaus2018} (4) \cite{Thalmann2014} (5) \cite{Thalmann2016} (6) \cite{Keppler2018} (7) \cite{Hashimoto2012} (8) \cite{Villenave2019} (9) \cite{deBoer2021}}
\label{tab: surface-profiles}
\end{table*}

\subsection{Extraction with \emph{diskmap}}

\begin{figure*}
\center
\includegraphics[width=0.99\textwidth]{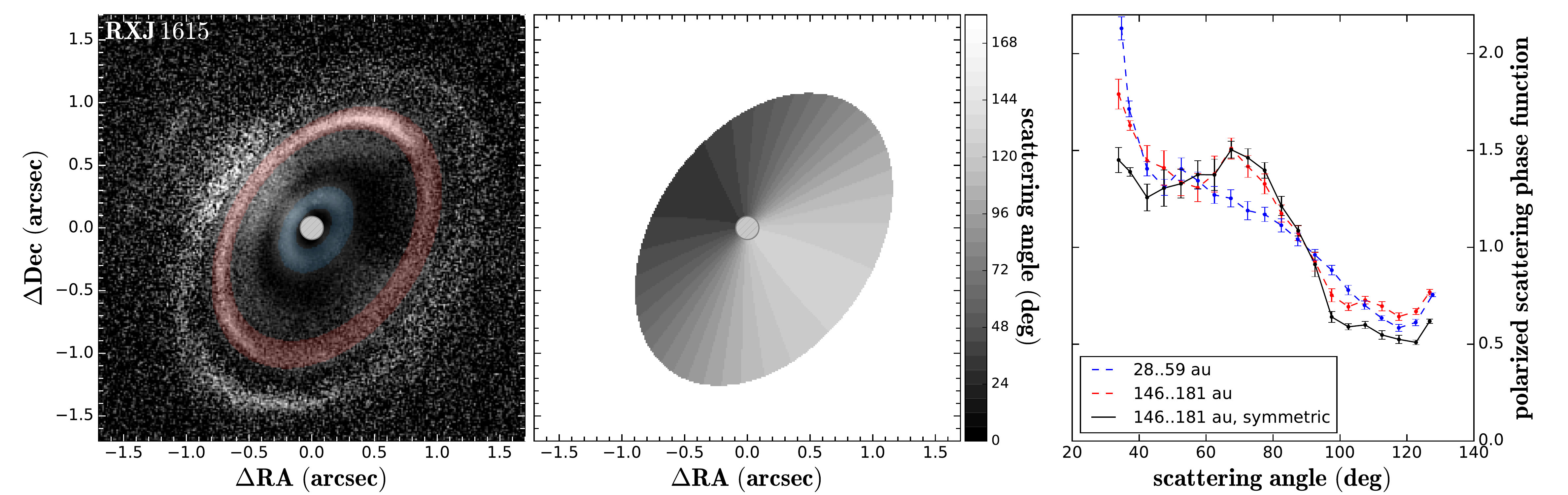} 
\caption{Extracted polarized scattered phase functions and extraction regions for the H-band data of the RXJ\,1615 system. \textit{Left:} H-band data of RXJ\,1615 scaled with the square of the distance from the central star. The blue and red transparent overlays highlight the two regions used for phase function extraction, i.e. the inner disk region close to the coronagraph and the bright outer ring. \textit{Middle:} Position dependent scattering angles calculated with the nominal surface height profile of the system. \textit{Right:} Extracted phase functions for the inner disk region (blue-dashed), the outer ring (red-dashed), and an azimuthal region of 180$^\circ$ centered on the north-western ansae of the outher ring (black-solid), which should be less affected by azimuthal shadowing. 
} 
\label{fig:rxj1615}
\end{figure*}

\begin{figure*}
\center
\includegraphics[width=0.99\textwidth]{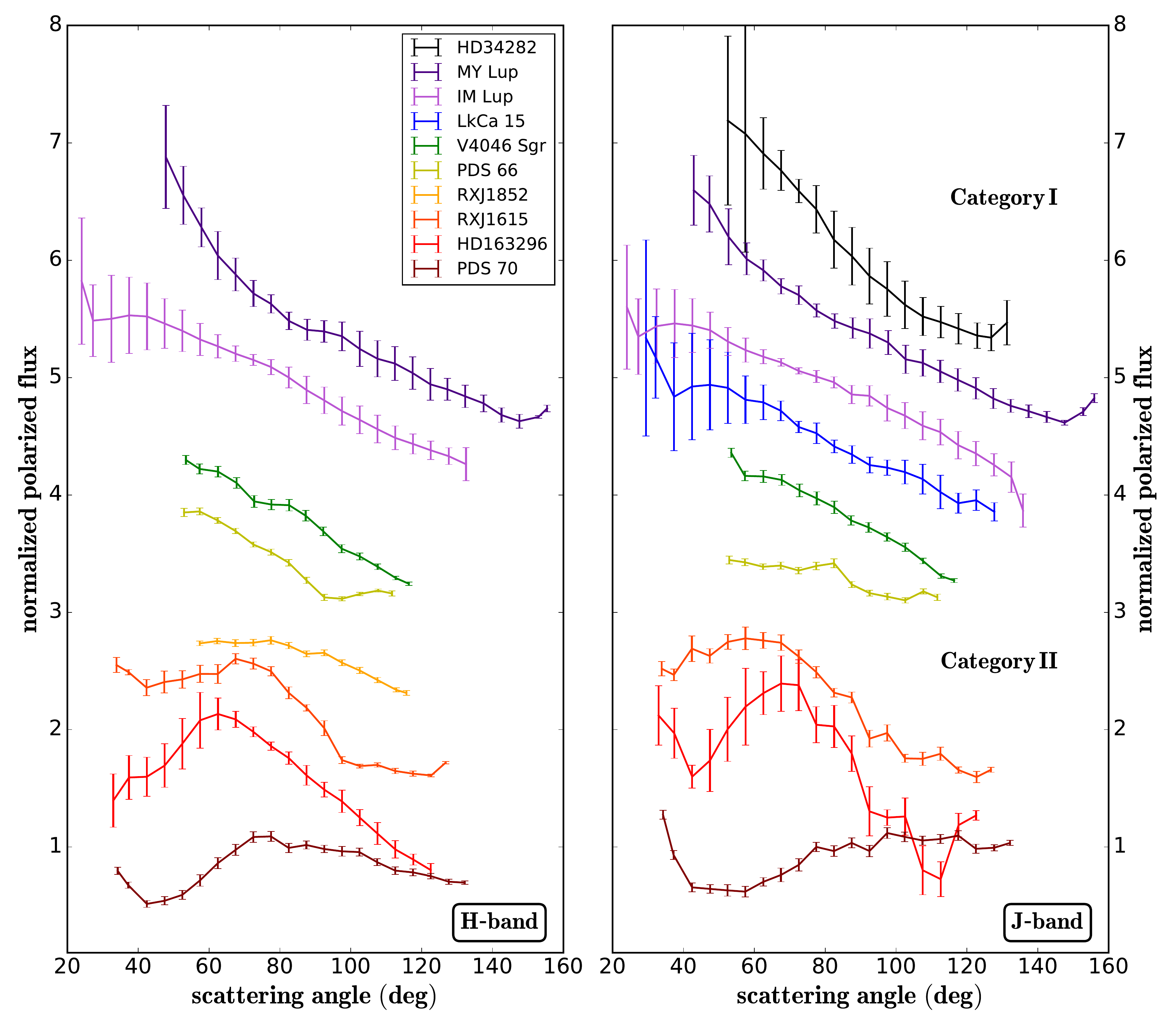} 
\caption{Final extracted polarized scattered light phase functions for all targets in our sample. We show the H-band data on the left and the J-band data on the right. All phase functions were normalized at scattering angles of 90$^\circ$ and had an arbitrary offset applied for better visibility. Colors indicate the same systems in both panels and the order of the phase functions from top to bottom is the same as indicated in the legend. Note that not for all targets both H and J-band were available.
} 
\label{fig:all phase functions}
\end{figure*}

For the extraction of the phase functions we use the publicly available \emph{diskmap} python package by \cite{Stolker2016}\footnote{https://diskmap.readthedocs.io}. While the package is described in detail in \cite{Stolker2016}, we give a brief summary here on the extraction steps. As described in the previous section the $\tau$=1 surface for all our system is described by a single power-law profile. Given this profile and the inclination of the disk \emph{diskmap} calculates for each pixel in the image the distance from the central star and subsequently the angle under which scattered light is received from this part of the disk. We show this for the RXJ\,1615 system in figure~\ref{fig:rxj1615} (middle panel).
The flux is corrected for the square-distance dependent illumination drop-off and is then extracted for each pixel, giving a single data point for the polarized scattering phase function. These data points are then placed in bins of scattering angles with a width of 5$^\circ$. For each bin the median scattering angle of all included pixels and the median flux is computed, giving the final data point for this angular bin. The standard deviation within each bin is used as a measure for the uncertainty of the phase function and captures effects due to the width of the bin as well as photometric uncertainties. \\
To include the uncertainty of the surface height profile we repeat the extraction for the "flat" and the "flared" extreme cases and calculate for each angle bin the flux difference. We consider this difference as the uncertainty of the phase function introduced by the uncertainty of the surface height profile. We quadratically combine this uncertainty with the standard deviation in each bin for the nominal extraction and consider the result the total uncertainty of the extracted phase function at each angle. \\
For each system we selected extraction regions centered on known bright ring features. Since the surface height profile is directly measured at the ring locations, this minimizes the introduced uncertainty while simultaneously selecting the region of the disk with the highest signal-to-noise ratio. If multiple rings are present then we separately extracted the phase function for each ring. In figure~\ref{fig:rxj1615} we show the two selected extraction regions for the RXJ\,1615 system. The extraction and individual phase functions for all systems are shown in appendix~\ref{appendix: all phase functions}.\\
In figure~\ref{fig:all phase functions} we show the final extracted phase function for all systems and photometric bands. We note that in figure~\ref{fig:all phase functions} we show the average phase function for the IM\,Lup and V\,4046\,Sgr systems instead of extractions for individual rings. For the HD\,163296 and RXJ\,1615 systems we show phase functions after correction for azimuthal shadowing as discussed in the following section.

\subsection{Effects of azimuthal shadowing}
\label{sec:shadowing}

\begin{figure}
\center
\includegraphics[width=0.49\textwidth]{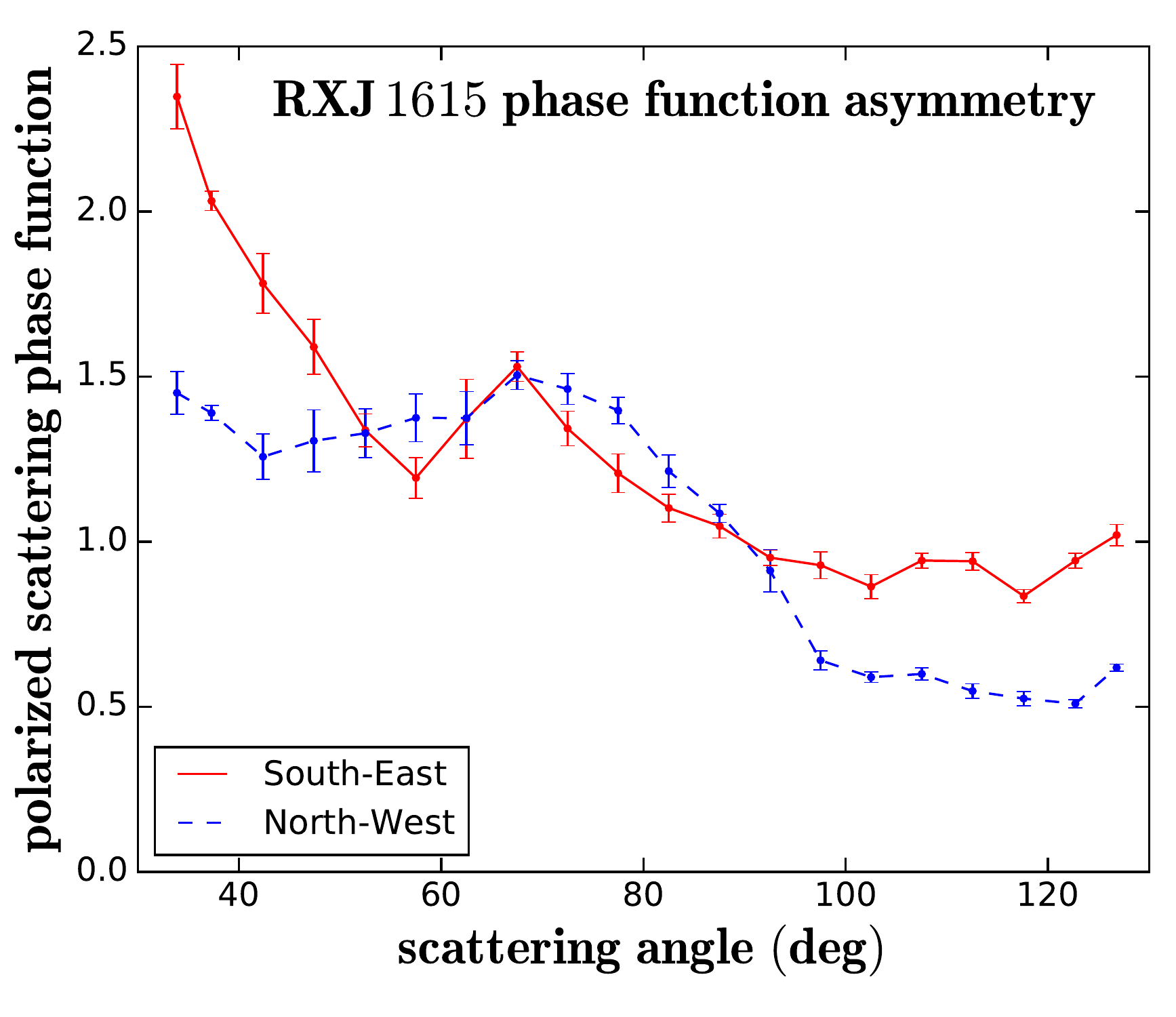} 
\caption{Polarized scattered phase functions of the RX\,1615 H-band data extracted from the brightest ring in the data set between projected separations of 146 and 181\,au. The blue-dashed phase function was extracted from an azimuthal region of 180$^\circ$ centered on the north-west ansae of the disk, while the red solid curve was extracted from a region centered on the south-east ansae. Both phase functions were normalized at scattering angles of 90$^\circ$. The differences may be explained by azimuthal shadowing of the extraction region due to an inner disk component, which strongly affected the south-western region. 
} 
\label{fig:asymmetry rxj1615}
\end{figure}

An aspect that may complicate the extraction of the phase function in some systems is azimuthal shadowing since it changes the brightness of disk regions due to reduced illumination, an effect that needs to be separated from the phase function. There are now a number of class II objects known for which shadows are observed in scattered light (see \citealt{Benisty2022} for an overview and detailed discussion). One of the most iconic systems is probably HD\,142527 (\citealt{Canovas2013,Avenhaus2014}) for which inner and outer disk are strongly misaligned (70$^\circ$, \citealt{Marino2015}), leading to two narrow shadows projected on the outer disk. Such narrow and well defined shadows are not of major concern for the extraction of the overall phase function as the small region affected by the shadows can simply be excluded. However, there as was shown for the HD\,139614 system by \cite{Muro-Arena2020}, small misalignments or warps in the disk can lead to very broad and somewhat diffuse shadowing, which covers in extreme cases an azimuthal range of more than 180$^\circ$. For such systems the problem is two fold. On the one hand the exclusion of large azimuthal regions of the disk from the extraction may severely limit the range of scattering angles for which the phase function can be extracted. On the other hand, these shadows are less well defined making it more difficult to decide which regions of the disk might be trusted for phase function extraction. \\
To estimate if the disks in our sample may be affected by broad shadowing effects we performed a simple analysis. If there is no shadowing present and the brightness distribution of disk structures is solely due to the dust scattering phase function\footnote{We imply here the assumptions that there are no azimuthal variations in the dust grain size distribution or composition, that the disks are not eccentric and there are no or only small azimuthal variations in surface density.}, then we would expect the disk surface brightness to be axis-symmetric relative to the disk minor axis. Thus for all our systems we flipped the disk images around the minor axis and then divided the original image by the flip image. This indicates the brightness ratio between the two "mirrored" sides of the disk. We show the axis-symmetric brightness ratio for all systems in the appendix in figure~\ref{fig:shadow gallery}. For most systems the deviation in brightness is smaller than a factor $\sim$2, with the notable exceptions of the RXJ\,1615 system (up to factor $\sim$4), the HD\,163296 system (up to factor $\sim$5) and the HD\,34282 system (up to factor $\sim$6). We thus consider that the RXJ\,1615 and the HD\,163296 systems maybe be affected by broad shadowing. This was in fact discussed for both systems in the literature by \cite{deBoer2016} and \cite{Muro-Arena2018}, respectively. In the case of HD\,34282 \cite{deBoer2021} discuss a possible spiral within the disk, thus in this case we may rather trace a genuine azimuthal asymmetry in the dust surface density rather than shadowing effects. \\
To give an indication how strongly these shadowing effects may affect the extracted phase function we performed two separate extractions for the RXJ\,1615 system choosing the bright ring between deprojected radii of 146\,au and 181\,au. One extraction only considered the brighter north-west side of the disk, while the second extraction only considered the fainter south-east side, i.e. the side of the ring more strongly affected by shadowing. The results are shown in figure~\ref{fig:asymmetry rxj1615} (both phase functions were normalized at scattering angles of 90$^\circ$). While the profiles somewhat match between 60$^\circ$ and 90$^\circ$ they deviate significantly at larger and smaller scattering angles. In particular the phase function extracted from the south-east half of the disk shows more relative flux in both ranges. If we consider that the shadow is centered on the south-east ansae, this might be explained by the stronger shadowing close to 90$^\circ$ scattering angles, or put in a different way, the disk may rise out of the shadow at regions seen under small and large scattering angles. \\
We note that there is no geometric reason why the disk should specifically be warped or misaligned around the minor axis. In practice there will in fact most likely be a deviation between the disk minor axis (defined by our arbitrary viewing geometry), and the warp or misalignment axis around which the disk tilts as a function of radial separation from the central star. However, if the axis of misalignment were closer to the disk major axis, then one would not expect a strong brightness asymmetry between the two disk ansae as seen in figure~\ref{fig:shadow gallery}. \cite{deBoer2016} also noted that the brightness asymmetry seemed to switch sides between subsequent ring structures in the disk, i.e. in the next outer (overall fainter) ring the south-east ansae is brighter than the north-west ansae indicating a continuous warp in the disk. If the disk near or far side were strongly affected by this effect then one may expect strong changes in brightness asymmetries between disk near and far side seen in subsequent rings. This does however not seem to be the case. We can thus conclude that the broad shadowing effect is likely centered at least near the disk ansae. This is further supported by the fact that the phase function extracted from the south-east side of the disk and shown in figure~\ref{fig:asymmetry rxj1615} increases the relative flux level toward both the forward and back-scattering sides of the disk. If the shadow were not centered close to the south-east ansae, then we would naively expect either a steeper phase function relative to the north-east at larger scattering angles or a flatter phase function at small scattering angles. \\
Given our analysis we thus consider the phase function extracted from the north west side of the disk in the RXJ\,1615 system to be minimally or in any case less affected by shadowing and use it for further analysis and comparison to other systems. This phase function is shown as black solid line in figure~\ref{fig:rxj1615} (named "symmetric" in the legend as this is the phase function of the point symmetric version of this disk relative to the minor axis). For the HD\,163296 system we then follow the same strategy and choose the north-west side of the disk for phase function extraction as it appears less affected by shadowing compared to the south-east.\\
For the HD\,34282 system the situation is different, as the detected asymmetry likely traces a genuine asymmetry in dust surface density profile, possibly due to spiral density waves in the disk gas. It is then not clear which regions may be best suited to extract an unbiased scattering phase function. We thus include in this case the full azimuthal range in the extraction and caution that a more detailed analysis of the system with dedicated radiative transfer modelling should be performed in the future to revise our preliminary results.

\subsection{Limb brightening}
\label{fig: limb brightening}

In addition to shadowing effects, the shape of the phase function may be influenced by the viewing geometry. In recent studies of optically thin debris disks \cite{Olofsson2020} and \cite{Engler2022} found that the relative flux extracted at $\sim$90$^\circ$ scattering angles, i.e. close to the disk ansae, may be enhanced compared to the intrinsic phase function of the present dust particles due to a higher column density along the line of sight. This effect is sometimes referred to as "limb brightening" (\citealt{Engler2022}).  
The systems in our sample are all at an earlier evolutionary stage before the gas dispersal in the disk and it is generally assumed that they are optically thick at near infrared wavelengths (\citealt{Chiang1997}). Thus we do not expect that the column density along the line of sight will change significantly for different scattering angles. However, due to the flaring surface height profiles of these systems we may instead expect an artificial increase in the flux ratio between the disk forward and back scattering sides, i.e. between small and large scattering angles. At small scattering angles (on the disk near side) the line of sight encompasses more of the illuminated disk surface than is the case for large scattering angles (on the disk far side). We discuss this effect in detail for the IM\,Lup system in Tazaki, Ginski \& Dominik (submitted). Using radiative transfer models we found in this study that the effect depends on the inclination, the local aspect ratio and flaring exponent of the disk surface height profile. For the specific case of the IM\,Lup system this may introduce a $\sim$25\% brightening for scattering angles smaller than $\sim$50$^\circ$. However, we caution that this strongly depends on the dust composition, i.e. for dust with high scattering albedos limb brightening may be much smaller or even insignificant because multiple scattering reduces the polarized flux at smaller scattering angles and starts to (at least partly) counteract this effect. Dedicated radiative transfer modelling is required to determine the influence of this effect for individual systems.\\
In a more general sense this limb brightening effect will typically not strongly alter the shape of the phase functions for optically thick disks, such as the ones discussed in our current study. Rather it will slightly increase the slope of the overall phase functions. We note that there may be one exception to this, which is the MY\,Lup system, which is seen under a particularly high inclination of 77$^\circ$. The extracted phase function in figure~\ref{fig:all phase functions} is strongly peaked toward small scattering angles. Based on the results in Tazaki, Ginski \& Dominik (submitted), we find it likely that the intrinsic phase function of dust particles in MY\,Lup has a significantly smaller slope, possibly with no up turn at small scattering angles. As we summarize in table~\ref{tab: surface-profiles} the remaining systems in our study have either a comparable inclination to IM\,Lup (this is the case for HD\,34282) or are seen under significantly smaller inclination. Thus, while the effect should be considered for future detailed modelling of individual systems, it will not strongly affect the general population level trends that we recover.   

\section{Results and Discussion}
\label{sec: discussion}

\subsection{Qualitative inference of dust properties}
\label{sec: dust properties}

Figure~\ref{fig:all phase functions} shows that the extracted polarization phase function is diverse in terms of their shapes. The shapes can be roughly divided into two categories: those that are monotonically increasing in polarized flux with decreasing scattering angle (category I: e.g., HD 34282, MY Lup, IM Lup, LkCa 15, V4046 Sgr) and those that turn around at a scattering angle of 60$^\circ$--80$^\circ$, go through a local minimum, and then increase again at the smallest scattering angles (category II: e.g., HD 163296, PDS 70).

\begin{figure*}[t]
\begin{center}
\includegraphics[width=0.49\linewidth]{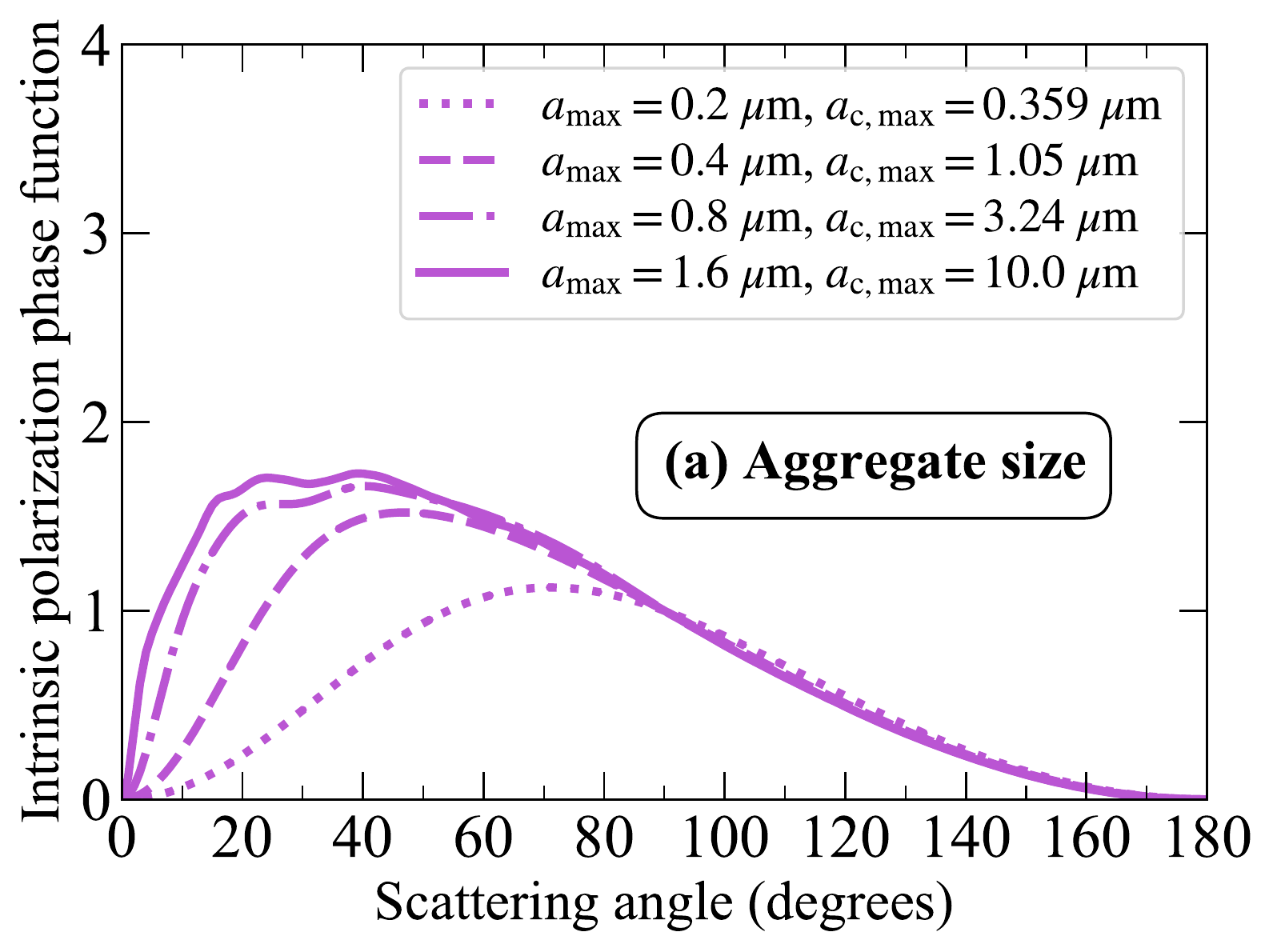}
\includegraphics[width=0.49\linewidth]{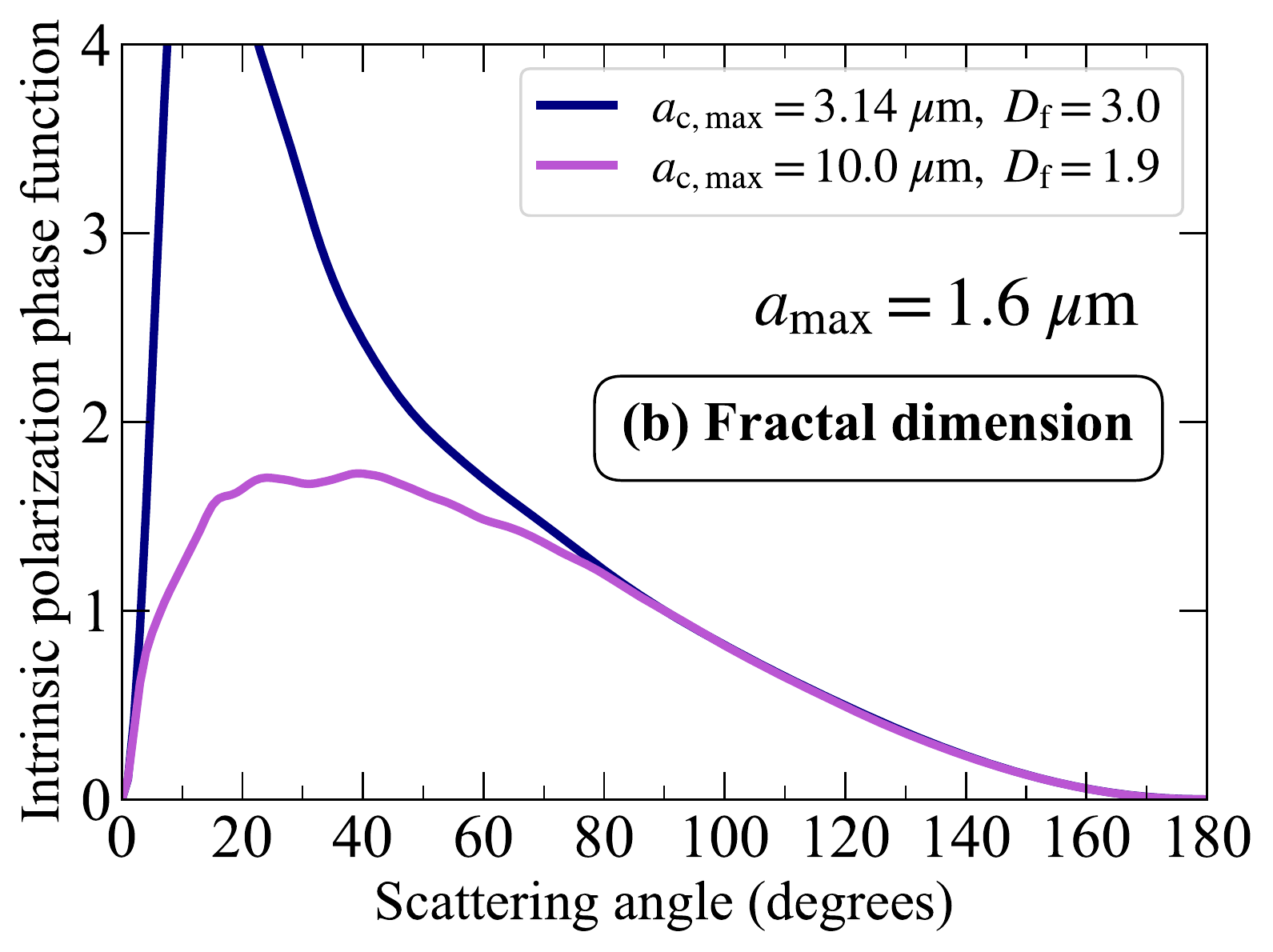}
\includegraphics[width=0.49\linewidth]{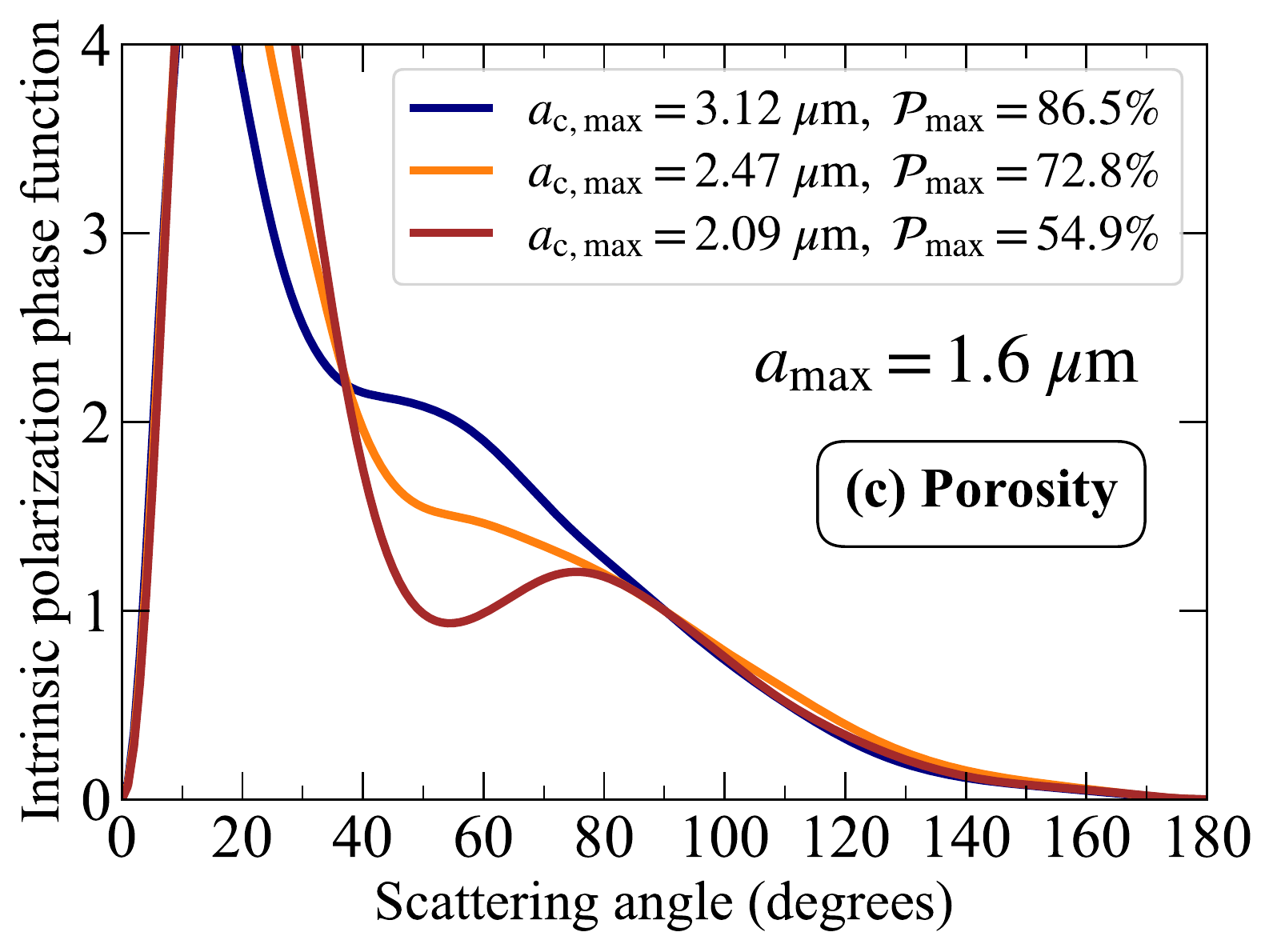}
\caption{Effect of aggregate size, fractal dimension, and porosity on the intrinsic polarization phase function. The phase functions are normalized to a scattering angle of 90$^\circ$. (a) 
The intrinsic functions for BCCA aggregates for various radii. The monomer radius is set as $\amon=0.1~\mu$m. (b)
The blue and violet lines represent the results for BPCA ($a_\mathrm{c, max}=3.14~\mu\mathrm{m},~D_\mathrm{f}=3.0$)  and BCCA ($a_\mathrm{c, max}=10.0~\mu\mathrm{m},~D_\mathrm{f}=1.9$) aggregates, respectively. In these computations we used $\amax=1.6~\mu$m and $\amon=0.1~\mu$m. 
(c) The blue, orange, and brown lines represent the results for BPCA ($a_\mathrm{c, max}=3.12~\mu\mathrm{m},~\mathcal{P}_\mathrm{max}=86.5\%$), BAM1 ($a_\mathrm{c, max}=2.47~\mu\mathrm{m},~\mathcal{P}_\mathrm{max}=72.8\%$), and BAM2 ($a_\mathrm{c, max}=2.09~\mu\mathrm{m},~\mathcal{P}_\mathrm{max}=54.9\%$) aggregates, respectively. In these computations we used $\amax=1.6~\mu$m and $\amon=0.4~\mu$m.}
\label{fig:model}
\end{center}
\end{figure*}

To illustrate the origins of these variations, we perform $T$-matrix light-scattering calculations for various dust aggregates \citep[][]{RT22} (see also Section \ref{sec:tmatrix}). The scattering matrix element ($-S_{12}$ in \citealt{Bohren83}) obtained by the simulations are summarized in figure \ref{fig:model}. There is a caveat when comparing the scattering matrix element and the observed phase function: a planet-forming disk is optically thick in the near infrared, and the scattering angle dependence of the observed polarization flux might have been affected by radiative transfer effects, such as multiple scattering within the disk surface and limb brightening (see a detailed discussion of these effects in Tazaki, Ginski \& Dominik, submitted). To distinguish it from the one extracted from an observed image, we will refer to the computed scattering matrix element as the {\it intrinsic polarization phase function}. While a three-dimensional radiative transfer calculation is necessary to determine the dust parameters more accurately from the extracted polarization phase function, it is possible to capture the general trend from the intrinsic phase function and infer the origins of the variations in shape shown in figure~\ref{fig:all phase functions}.

Figure \ref{fig:model} (a) illustrates how the intrinsic polarization phase function of aggregates varies with their size. The scattering angle dependency of the polarized intensity is the result of the dependence of the two quantities overlapping: total-intensity phase function and the degree of linear polarization. When the aggregates are small compared to the wavelength, the scattered light distribution will be close to isotropic in total intensity, but the degree of polarization approaches 0 in the forward scattering direction. As a result, the intrinsic polarization phase function turns around at an angle of scattering around 90$^\circ$ (for the case of pure Rayleigh scattering, the intrinsic polarization function is proportional to $1-\cos^2\theta$, see \citet{Bohren83}). As the aggregates become larger, forward scattering in total intensity develops and compensates for the decrease in the degree of polarization. Consequently, the turn-around position shifts to the small-angle side. 

Except for RXJ1852, none of the observed phase functions presented in figure~\ref{fig:all phase functions} exhibits the Rayleigh-scattering-like profile. This indicates that the aggregates have grown to at least micron sizes in those disks. For RXJ1852, the function appears to have a peak in polarized flux around a scattering angle of 80$^\circ$, and the profile may be consistent with the presence of small, submicron-sized aggregates. Since the presence of such small particles makes the disk scattered light bluish (\citealt{Tazaki2019}), future multi-color observations would be useful to draw a conclusion.

The shape diversity in polarization phase functions could be resulting from a diversity of the structure and porosity of micron-sized aggregates. 
First of all, we focus on category I (the top six curves in the $J$ band); all show monotonically increasing polarizing flux with decreasing scattering angle. 
In particular, the curves for V4046 Sgr, LkCa 15, and IM Lup seem to have approximately constant slopes, while the slope of the curves for MY Lup and HD 34282 become steeper at scattering angles below 80$^\circ$. 
Figure \ref{fig:model} (b) demonstrates that aggregates with different fractal dimensions can explain these differences. Aggregates with a low fractal dimension (around 1.9) exhibit nearly constant slopes except for scattering angles below 30$^\circ$, whereas aggregates with a high fractal dimension (around 3.0) show a similar slope to fractal aggregates in the large scattering angle region, but the slope becomes steeper in the small scattering angle region.
Therefore, the approximately constant slope observed in V4046 Sgr, LkCa 15, and IM Lup may be explained by fractal aggregates. In fact, the detailed radiative transfer calculations by Tazaki, Ginski and Dominik (submitted) showed that fractal aggregates best explains the observed phase function of the IM Lup disk. On the other hand, the phase functions for MY Lup and HD 34282  point to the presence of aggregates with a high fractal dimension, although the porosity is still high ($\sim87\%$), unless the limb brightening is responsible for the steepening of the slope. Therefore, these disks might contain highly porous aggregates, although a full radiative transfer modeling is needed to determine the fractal dimension.

Category II exhibits peculiar phase functions (e.g., RXJ 1615, HD 163296, PDS 70 in figure~\ref{fig:all phase functions}). Firstly, there is a turn-around at scattering angles of 60$^\circ$--80$^\circ$, and secondly the polarized flux increases again at the smallest scattering angles. The similar trend has also been reported in the disk around HD 100546 \citep{Stolker2016}. The effects of radiative transfer within the disk surface would not account for this trends as long as the disk structure is axisymmetric and is uniformly illuminated by the central star. For example, multiple scattering tends to decrease the polarization flux on the forward scattering side, but its dependency is monotonic and would not explain the re-rise at the smallest scattering angles. If this is the case, the tendency has to be attributed to the intrinsic properties of dust particles. However, the high porosity aggregates that are thought to explain category I do not show such a tendency, as already shown in figure~\ref{fig:model} (a) and (b). This means that the dust particles in these system are different from the ones in category I. 

One possibility is low-porosity aggregates. Figure~\ref{fig:model}(c) shows that a turn-around at scattering angles around 80$^\circ$ and a re-rise in polarization flux at small scattering angles appear simultaneously when the porosity is as low as $\sim55\%$. Since this feature is not prominent for a higher porosity, it seems to be triggered by an increased contribution of monomer-monomer electromagnetic interaction as the monomers get packed closely for lower porosity. Therefore, the phase functions for the disks around HD 163296, PDS 70, and perhaps RXJ1615, might be explained by low porosity aggregates.
Low porosity aggregates tend to show a reddish polarized intensity color (\citealt{Tazaki2019}). For RXJ\,1615 \cite{Avenhaus2018} found that J/H=0.78$\pm$0.42 thus indeed the disk appears red, i.e. is brighter in the H-band relative to the central star than in the J-band. We repeated an analog measurement to that described in \cite{Avenhaus2018} for the HD\,163296 and the PDS\,70 systems. We found J/H=0.75$\pm$0.11 and J/H=0.94$\pm$0.21 for HD\,163296 and PDS\,70, respectively. Thus the polarized intensity colors for all three systems are consistent with the presence of low porosity aggregates in the surface layers of the disk (although we caution that in the cases of RXJ\,1615 and PDS\,70 the error bars are large on the color measurement).

Low porosity aggregates have relative small area-to-mass ratios (i.e.g, weak dynamical coupling with gas), making them prone to settling on the disk midplane. Efficient vertical mixing would be necessary to keep these aggregates in the disk surface layers \citep[e.g.,][]{Mulders13, RT21c}. Interestingly, the disks around HD 163296, PDS 70, and HD 100546 (studied in \citealt{Stolker2016}) have been suggested to host a planet in the disk (\citealt{Teague2018, Keppler2018, Haffert2019, Quanz2013, Currie2014, Casassus2019}). The presence of planets has been suggested to influence vertical mixing of dust particles in disks by meridional circulation\citep{Bi21, Binkert21, Szulagyi22}. Therefore, we speculate that the presence of planets might indirectly affect the dust properties at the disk surface, which might in turn explain the distinct phase functions from the others.

\subsection{Trends with system properties}

\begin{figure*}
\center
\includegraphics[width=0.99\textwidth]{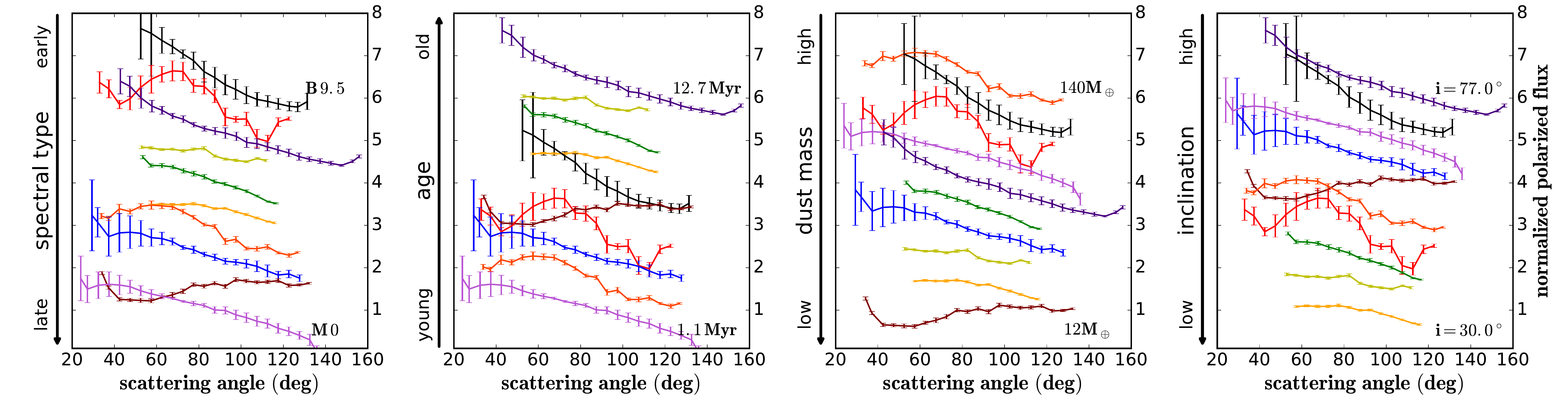} 
\caption{Phase functions for all 10 target systems ordered by various system parameters. We show the J-band data for all systems but one since it is more complete. For the RXJ\,1852 system we show the H-band data as there are no J-band observations of this system. The color code is the same as in figure~\ref{fig:all phase functions} for all systems. We indicate for each panel on the left y-axis the system parameter by which the phase functions were sorted and indicate the extreme values of these parameters in the plot.
} 
\label{fig: phase function sequences}
\end{figure*}

As discussed in section~\ref{sec: dust properties} we find an empirical dichotomy in the shape of the phase functions indicating different dust aggregate porosities. In order to find possible trends with basic system parameters we plot the phase functions of all systems in order of stellar spectral type, system age, disk dust mass and disk inclination in figure~\ref{fig: phase function sequences}. For the stellar spectral type and the disk dust mass we do not see any correlation between the two different categories of phase functions. For the system age we notice that the three disks with lower dust porosities are among the younger sources in our sample, with ages of 5.4\,Myr for PDS\,70 (\citealt{Mueller2018}), 5.1\,Myr for HD163296 (\citealt{Alecian2013}) and 1.4\,Myr for RXJ\,1615 (\citealt{Wahhaj2010}). However the presumably youngest source in our sample, IM\,Lup (1.1\,Myr, \citealt{Avenhaus2018}) is again part of the higher porosity category. We stress that our sample is small and that individual system ages are inherently uncertain, thus an expanded study with a large number of systems will be needed to confirm if such a trend indeed exists. \\
In the rightmost panel of figure~\ref{fig: phase function sequences} we order the phase functions by disk inclination. Here we find that the two phase functions with a monotonous slope and strong forward scattering peak were extracted from the two systems with the highest inclination in our study, i.e. MY\,Lup (77$^\circ$) and HD\,34282 (57$^\circ$). Conversely the two phase functions for the systems with the lowest inclination show a monotonous and shallow slope with no indication for a similar forward scattering peak (although we note that in these cases the smallest scattering angles could not be sampled). This trend with inclination may well correspond to the limb brightening effect discussed in section~\ref{fig: limb brightening}.

\subsection{Multi-ringed systems}

For three systems with multiple ring-like features we were able to extract the polarized phase function at multiple separations: RXJ\,1615, IM\,Lup and V\,4046\,Sgr. \\
For \textit{RXJ\,1615} we extracted the phase function from the innermost resolved ring between 28\,au and 59\,au as well as the brightest full ring at a radial separation between 146\,au and 181\,au. The resulting extractions in the H-band are shown in figure~\ref{fig:rxj1615}. For the comparison we consider the extraction of the outer ring which was corrected for azimuthal shadowing as discussed in section~\ref{sec:shadowing} (the black, solid curve in figure~\ref{fig:rxj1615}). The two phase functions show a very different shape. While the outer disk shows the previously mentioned peak between 60$^\circ$ and 80$^\circ$, the phase function of the inner disk zone is well described by a single slope for angles between 40$^\circ$ and 120$^\circ$, but shows strong peaks at larger and smaller angles. Both of them seem to favor compact aggregates that have relatively high fractal dimensions, e.g., the BPCA aggregates shown in figure~\ref{fig:model} (b). In order to explain the dip, the porosity needs to be relatively low, e.g., $\mathcal{P}_\mathrm{max}\sim55~\%$, at least for the outer region (figure~\ref{fig:model} c). Since the phase functions are very different from each other, the inner and outer disk surfaces are likely dominated by different types of compact dust aggregates. 
As we already discuss in the previous section, the presence of low porosity aggregates in the upper disk atmosphere may indicate the presence of a perturber that leads to more efficient vertical mixing. This fits also well with our discussion in section~\ref{sec:shadowing} which indicates that there is a warp present in the outer disk of the RXJ\,1615 system, consistent with previous results by \cite{deBoer2016}.\\
We discuss the \textit{IM\,Lup} system in great detail in Tazaki, Ginski \& Dominik (submitted). As a brief summary we note that the two innermost disk zones between 70\,au and 110\,au and between 130\,au and 170\,au are well consistent with each other (see figure~\ref{fig:imlup}). The outermost extraction zone between 217\,au and 257\,au shows a much shallower slope toward small scattering angles (but also has intrinsically much larger uncertainties due to the less well defined surface height profile in the outer disk, see figure~\ref{fig:angle gallery}). As we argue in Tazaki, Ginski \& Dominik (submitted), this may simply be an indication that the outer disk region is not well described anymore by the same power-law profile as the inner regions due to decreasing surface density and thus decreasing optical depth. \\
The \textit{V\,4046\,Sgr} system has arguably the most well defined dual ring structure of all disks in this study and shows no indication of significant azimuthal shadowing. We extracted the phase function from the inner region between 11\,au and 19\,au as well as from the outer ring between 23\,au and 34\,au (see figure~\ref{fig:v4046sgr}). We find that both phase functions are well consistent with each other for angles larger than $\sim$80$^\circ$, while the inner disk shows a slightly (but significantly) smaller slope for smaller scattering angles. It is unclear if this slight difference is due to the intrinsic phase functions in the inner and outer ring (and thus indicates slightly different aggregate properties) or if it can be attributed to observational effects such as limb brightening. Indeed, due to the steeper slope in the outer ring of the flared disk surface we would expect a slightly stronger limb brightening effect for the outer ring, which may then explain the small deviations between the two observed phase functions. 

\section{Summary}
\label{sec: summary}

We measure for the first time the polarized scattering phase functions of 10 young planet forming disks observed in the near infrared. We find that even though the geometry of these disks is complex, phase functions with meaningful uncertainties can be extracted. While detailed radiative transfer models for individual systems are required to disentangle observational effects such as limb brightening or azimuthal shadowing from the intrinsic phase function of the dust particles, we can still infer some general trends from the extracted phase functions.

\begin{enumerate}

\item We find empirically two distinct categories of phase functions. Category I has a monotonous slope, while category II displays a local maximum between 60$^\circ$  to 80$^\circ$.

\item Disks in category I have phase functions consistent with micron-sized, high porosity aggregates, while disks in category II require micron-sized, low porosity aggregates to explain their phase function. 

\item Category II disks appear consistent with red polarized intensity colors between the J/H band, as predicted for micron-sized, low porosity aggregates.

\item While we do not find general correlations with basic system parameters for the two phase function categories we do note that category II disks include the HD\,163296 and the PDS\,70 systems, both of which host embedded planets. Furthermore the literature data of the HD\,100546 system, which is also suggested to host planets is consistent with phase function category II.

\item If the presence of low porosity aggregates is an indication for the presence of embedded planets, then this may indicate that the RXJ\,1615 system, which also belongs to the category II disks, hosts an embedded planet similar to the cases of PDS\,70 and HD\,163296

\end{enumerate}

As further near infrared observations of young disks become available it will be most interesting to repeat the extraction performed for the small sample in this study to investigate if the two tentative categories that we identify are indeed present in the larger population.

\acknowledgments

C.G. dedicates this work to the memory of Gisela Else M\"{a}der. Her many years of selfless support enabled his past and present scientific work.  
C.G. acknowledges funding from the Netherlands Organisation for Scientific Research (NWO) TOP-1 grant as part
of the research program “Herbig Ae/Be stars, Rosetta stones for understanding
the formation of planetary systems”, project number 614.001.552.
R.T. acknowledges the JSPS overseas research fellowship. 
R.T. also thank Daniel Mackowski for making the MSTM codes publicly available. R.T. awould also like to thank Bruce Draine for the availability of particle data of BA, BAM1, and BAM2 and Yasuhiko Okada for providing a generation code for BCCA.
SPHERE is an instrument designed and built by a consortium
consisting of IPAG (Grenoble, France), MPIA (Heidelberg, Germany), LAM (Marseille, France), LESIA (Paris, France), Laboratoire Lagrange (Nice, France), INAF - Osservatorio di Padova (Italy), Observatoire de
Gen\`{e}ve (Switzerland), ETH Zurich (Switzerland), NOVA (Netherlands), ONERA
(France), and ASTRON (The Netherlands) in collaboration with ESO.
SPHERE was funded by ESO, with additional contributions from CNRS
(France), MPIA (Germany), INAF (Italy), FINES (Switzerland), and NOVA
(The Netherlands). SPHERE also received funding from the European Commission
Sixth and Seventh Framework Programmes as part of the Optical Infrared
Coordination Network for Astronomy (OPTICON) under grant number RII3-Ct2004-001566
for FP6 (2004-2008), grant number 226604 for FP7 (2009-2012),
and grant number 312430 for FP7 (2013-2016).
This research has used the SIMBAD database, operated at CDS, Strasbourg, France \citep{Wenger2000}. 
We used the \emph{Python} programming language\footnote{Python Software Foundation, \url{https://www.python.org/}}, especially the \emph{SciPy} \citep{2020SciPy-NMeth}, \emph{NumPy} \citep{oliphant2006guide}, \emph{Matplotlib} \citep{Matplotlib} and \emph{astropy} \citep{astropy_1,astropy_2} packages.
We thank the writers of these software packages for making their work available to the astronomical community.

\newpage

%

\vspace{2mm}
\facilities{VLT(SPHERE)}
\software{\texttt{MSTM v3.0} \citep{Mackowski11}}

\bibliography{MyBibFMe}{}
\bibliographystyle{aasjournal}

\begin{appendix}

\section{Summary of observation setups and conditions}
\label{app: observations}

\begin{table*}
 \centering
 \caption{Observing dates, instrument setup and weather conditions for all systems in our study.}
  \begin{tabular}{lccccccc}
  \hline 
 Target  	& Date 		& Filter 		& DIT\,[s] 	&  \# frames 	& Seeing\,[arcsec]	& $\tau_0$\,[ms] & ESO ID 	\\
 \hline
RXJ\,1615.3-3255    & 14-03-2016 & BB\_H & 64 & 80 & 1.1     & 2.6 &  096.C-0523(A) \\
                    & 14-03-2016 & BB\_J & 64 & 48 &  1.1   & 2.2 & 096.C-0523(A) \\
HD\,163296          & 26-05-2016 & BB\_H & 8 (16) & 64 (64) & 1.0    & 2.8 & 097.C-0523(A) \\
                    & 26-05-2016 & BB\_J & 16 & 200 & 0.8    & 2.0 & 097.C-0523(A) \\
IM\,Lup             & 13-03-2016 & BB\_H & 64 & 48 & 1.3     & 3.4 & 096.C-0523(A) \\
                    & 11-03-2016 & BB\_J & 64 & 56 & 0.9    & 1.5 & 096.C-0523(A) \\
LkCa\,15            & 19-12-2015 & BB\_J & 32 & 120 & 0.7   & 2.4 & 096.C-0248(A) \\
PDS\,66             & 15-03-2016 & BB\_H & 64 & 56 & 0.9     & 2.8 & 096.C-0523(A) \\
                    & 14-03-2016 & BB\_J & 64 & 48 & 1.1    & 2.5 & 096.C-0523(A) \\ 
PDS\,70             & 09-08-2019 & BB\_H & 64 & 36 &     1.5 & 2.1 & 0102.C-0916(B) \\
                    & 26-03-2016 & BB\_J & 64 & 52 & 1.9    & 1.3 & 096.C-0333(A) \\
2MASSJ18521730-3700119         & 15-05-2017 & BB\_H & 64 & 12 &   1.2   & - & 099.C-0147(B) \\
V\,4046\,Sgr        & 13-03-2016 & BB\_H & 64 & 48 & 1.0    & 2.3 & 096.C-0523(A) \\
                    & 12-03-2016 & BB\_J & 64 & 48 & 1.5    & 1.7 & 096.C-0523(A) \\                    
HD\,34282           & 19-12-2015 & BB\_J & 64 & 88 &   0.6  & 3.0 & 096.C-0248(A) \\
MY\,Lup             & 15-03-2016 & BB\_H & 64 & 40 & 0.7    & 3.7 & 096.C-0523(A) \\
                    & 15-03-2016 & BB\_J & 64 & 35 & 0.9    & 2.4 & 096.C-0523(A) \\
 
\hline\end{tabular}
\label{tab: observations}
\end{table*}

\section{Phase function extraction of all systems}
\label{appendix: all phase functions}

\begin{figure*}
\center
\includegraphics[width=0.79\textwidth]{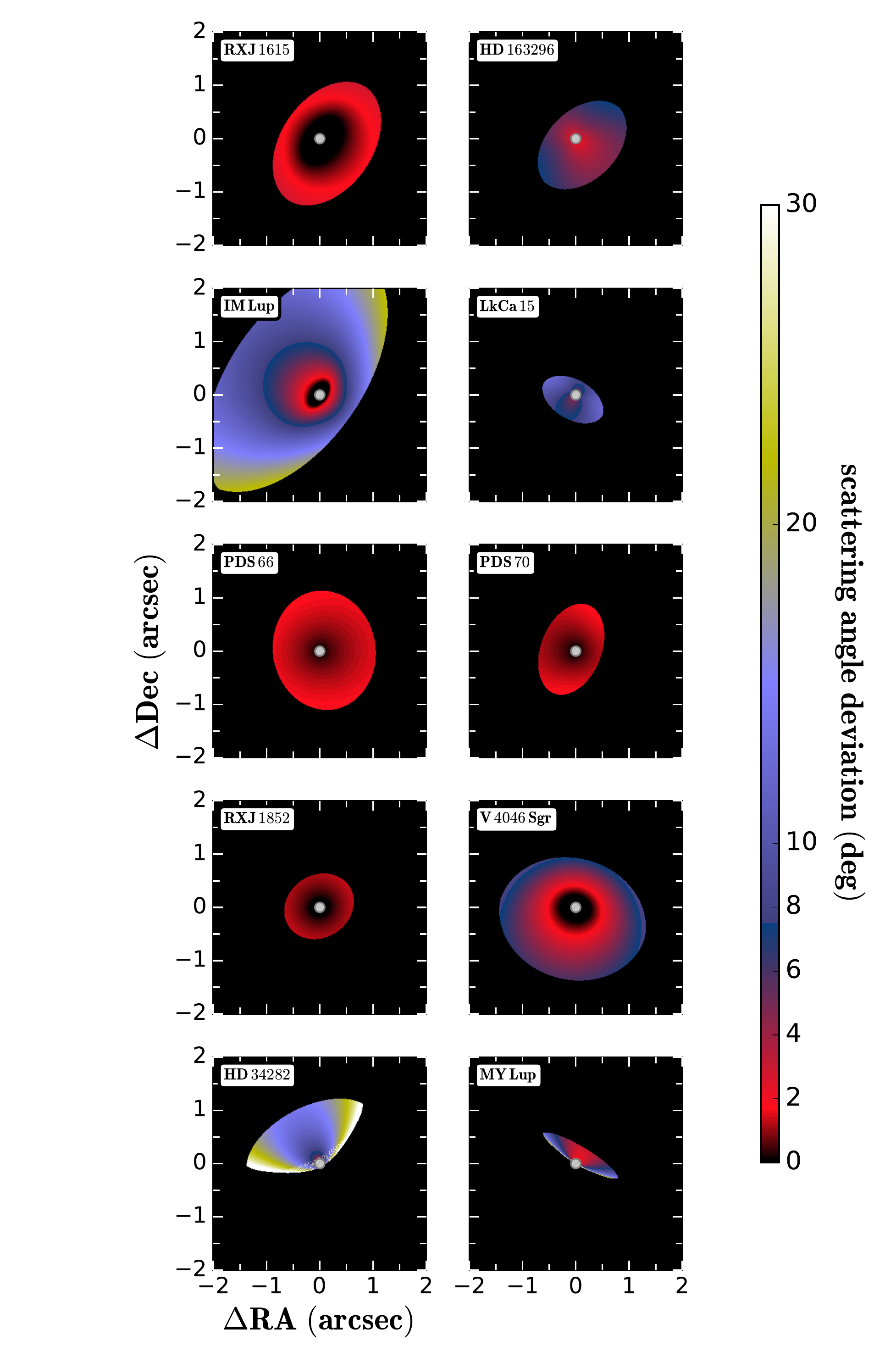} 
\caption{Same as the bottom panel of figure~\ref{fig:surface profile rxj1615} but for all disks in our sample. Shown are the maximum deviations in scattering angle based on the flared and the flat surface height profiles.
} 
\label{fig:angle gallery}
\end{figure*}

\begin{figure*}
\center
\includegraphics[width=0.99\textwidth]{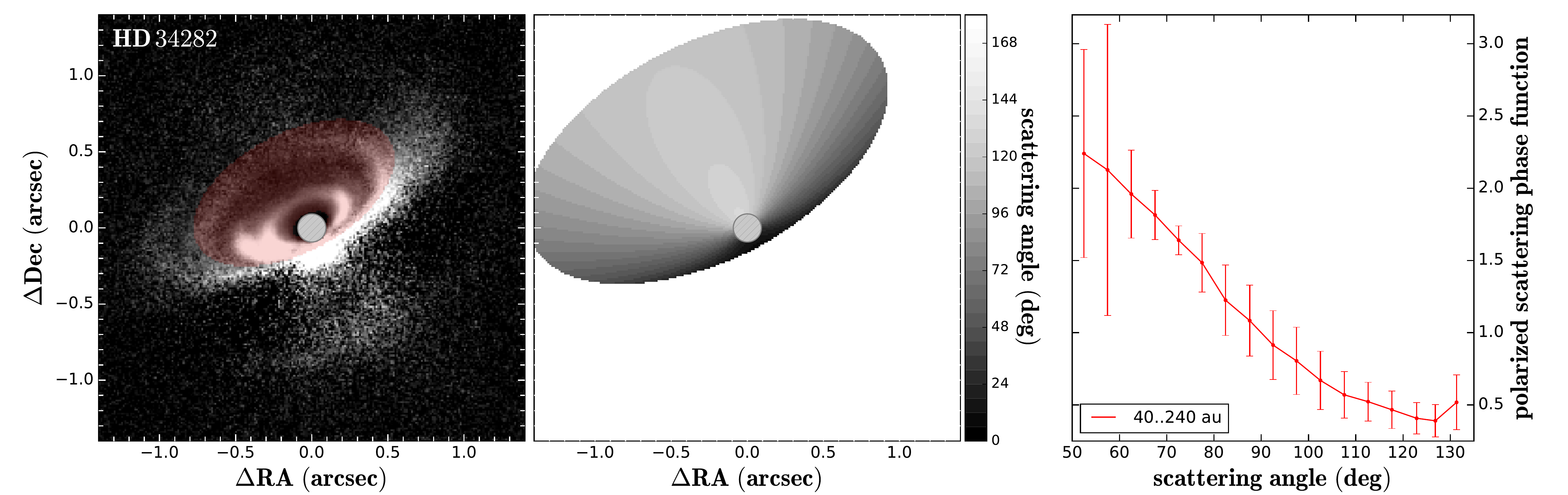} 
\caption{Extracted polarized scattered phase functions and extraction regions for the J-band data of the HD\,34282 system. Panels are analog to figure~\ref{fig:rxj1615}.
} 
\label{fig:hd34282}

\end{figure*}

\begin{figure*}
\center
\includegraphics[width=0.99\textwidth]{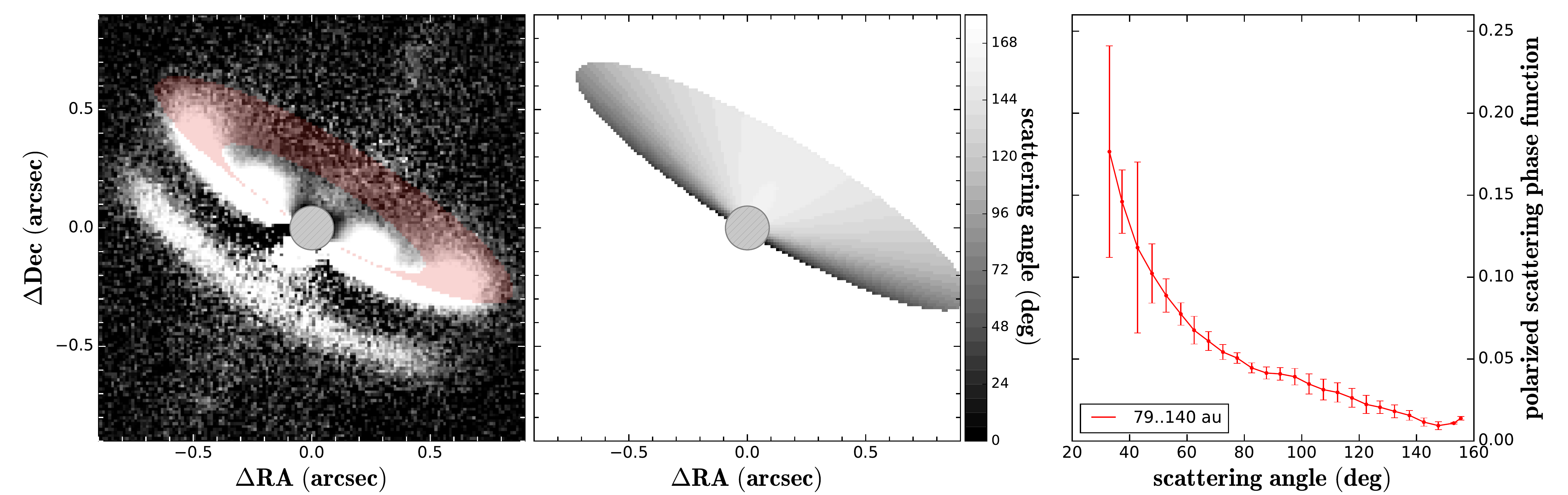} 
\caption{Extracted polarized scattered phase functions and extraction regions for the H-band data of the MY\,Lup system. Panels are analog to figure~\ref{fig:rxj1615}.
} 
\label{fig:mylup}
\end{figure*}

\begin{figure*}
\center
\includegraphics[width=0.99\textwidth]{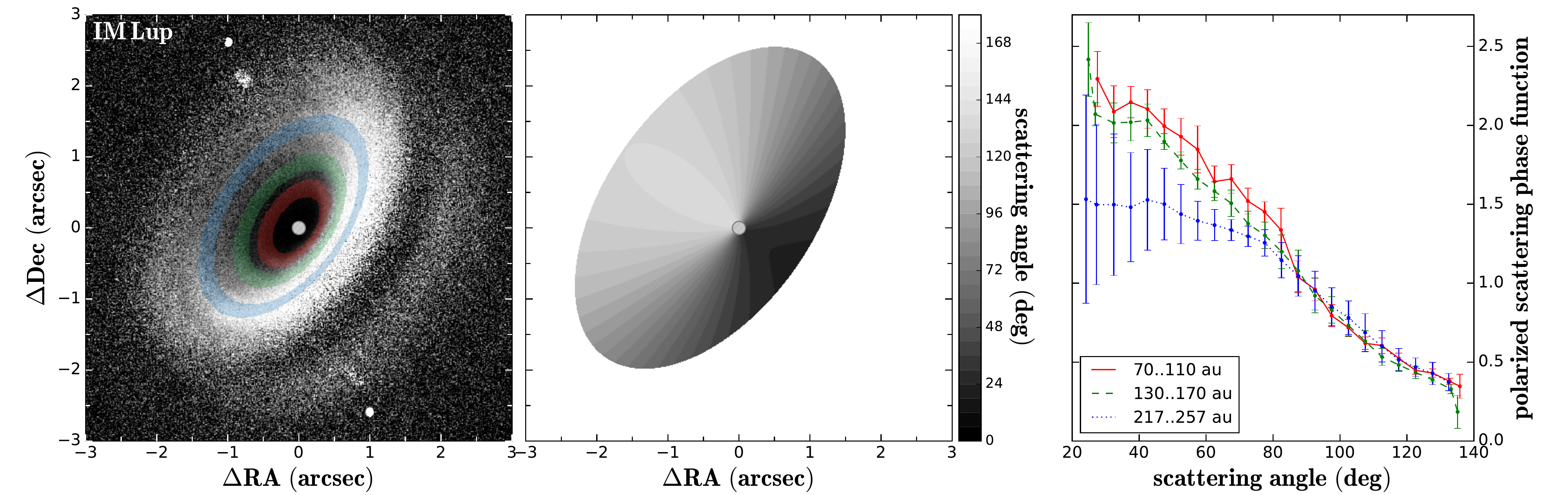} 
\caption{Extracted polarized scattered phase functions and extraction regions for the H-band data of the IM\,Lup system. Panels are analog to figure~\ref{fig:rxj1615}. We note that this figure is identical to figure~1 in Tazaki et al. (submitted).
} 
\label{fig:imlup}
\end{figure*}

\begin{figure*}
\center
\includegraphics[width=0.99\textwidth]{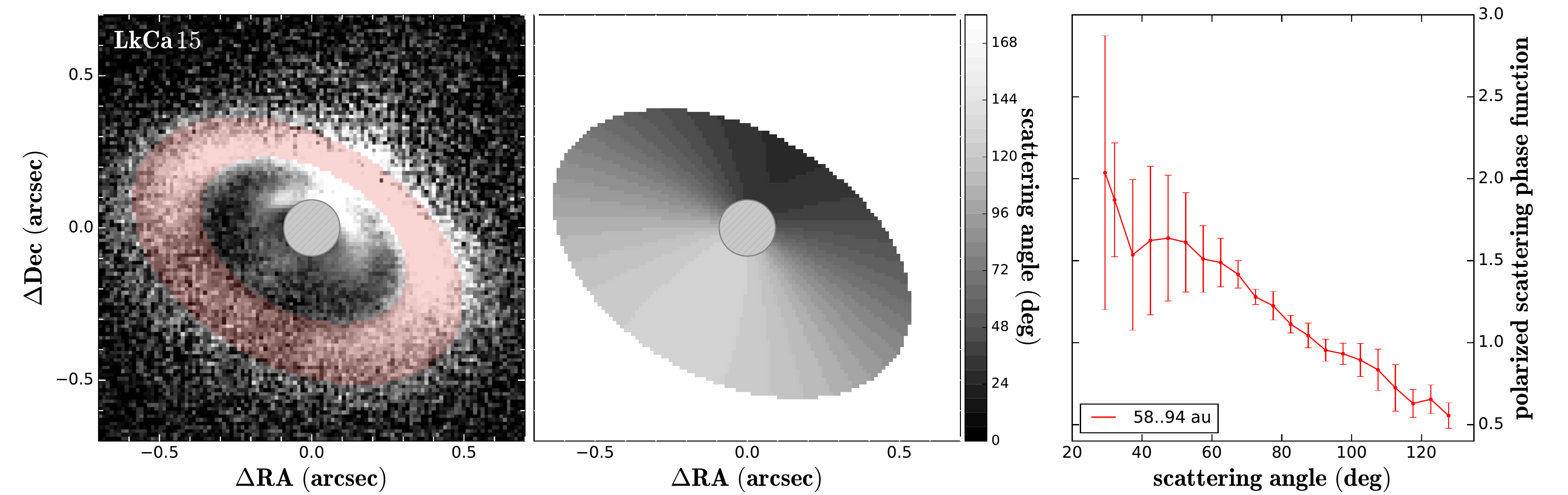} 
\caption{Extracted polarized scattered phase functions and extraction regions for the J-band data of the LkCa\,15 system. Panels are analog to figure~\ref{fig:rxj1615}.
} 
\label{fig:lkca15}
\end{figure*}

\begin{figure*}
\center
\includegraphics[width=0.99\textwidth]{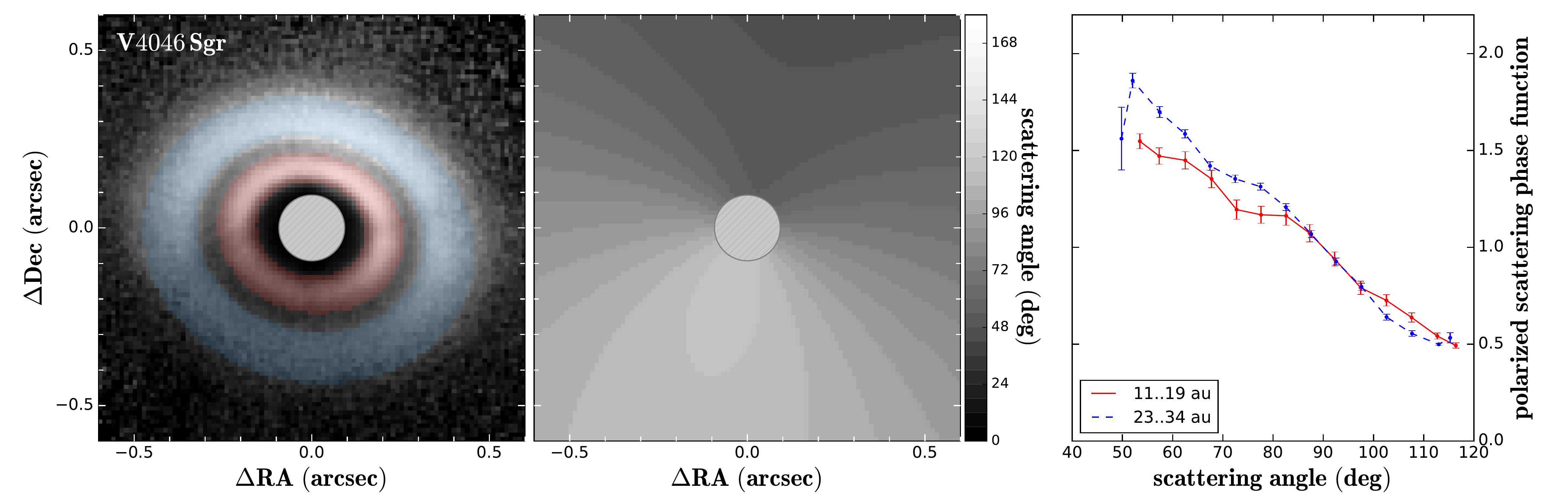} 
\caption{Extracted polarized scattered phase functions and extraction regions for the H-band data of the V\,4046\,Sgr system. Panels are analog to figure~\ref{fig:rxj1615}.
} 
\label{fig:v4046sgr}
\end{figure*}

\begin{figure*}
\center
\includegraphics[width=0.99\textwidth]{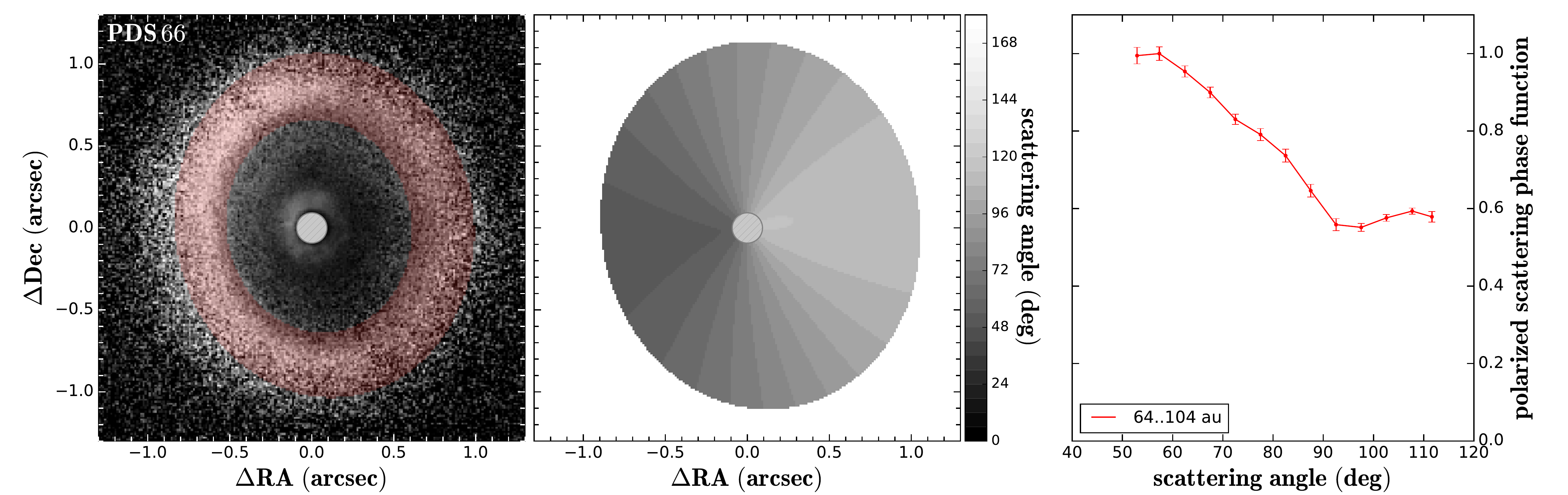} 
\caption{Extracted polarized scattered phase functions and extraction regions for the H-band data of the PDS\,66 system. Panels are analog to figure~\ref{fig:rxj1615}.
} 
\label{fig:pds66}
\end{figure*}

\begin{figure*}
\center
\includegraphics[width=0.99\textwidth]{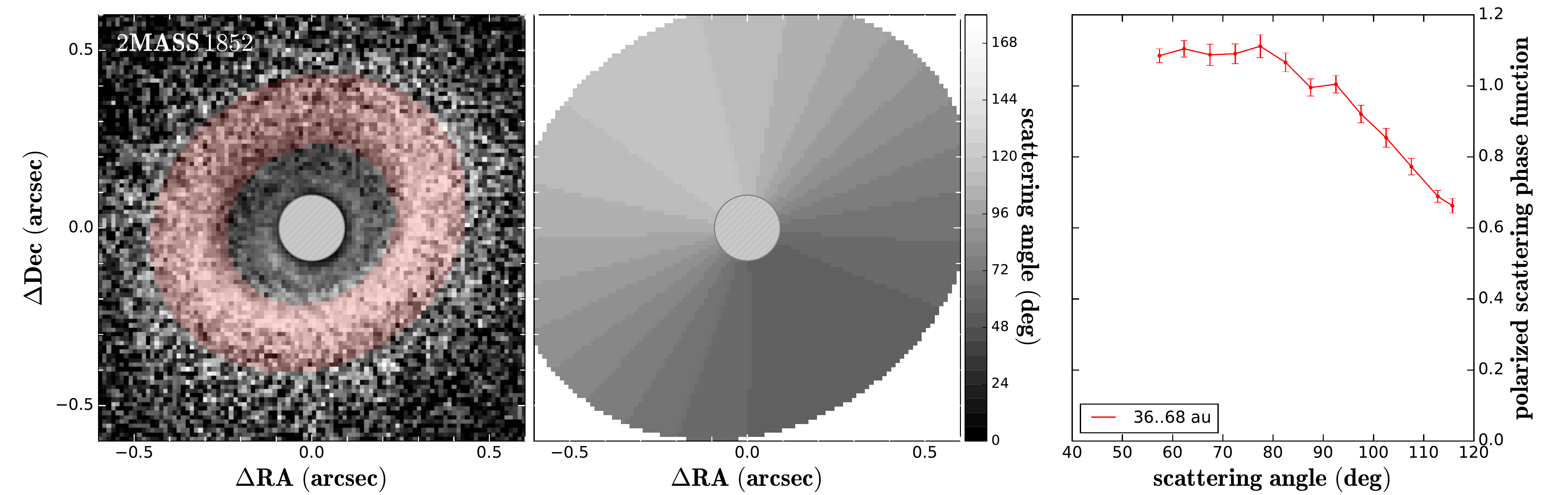} 
\caption{Extracted polarized scattered phase functions and extraction regions for the H-band data of the RXJ\,1852 system. Panels are analog to figure~\ref{fig:rxj1615}.
} 
\label{fig:rxj1852}
\end{figure*}

\begin{figure*}
\center
\includegraphics[width=0.99\textwidth]{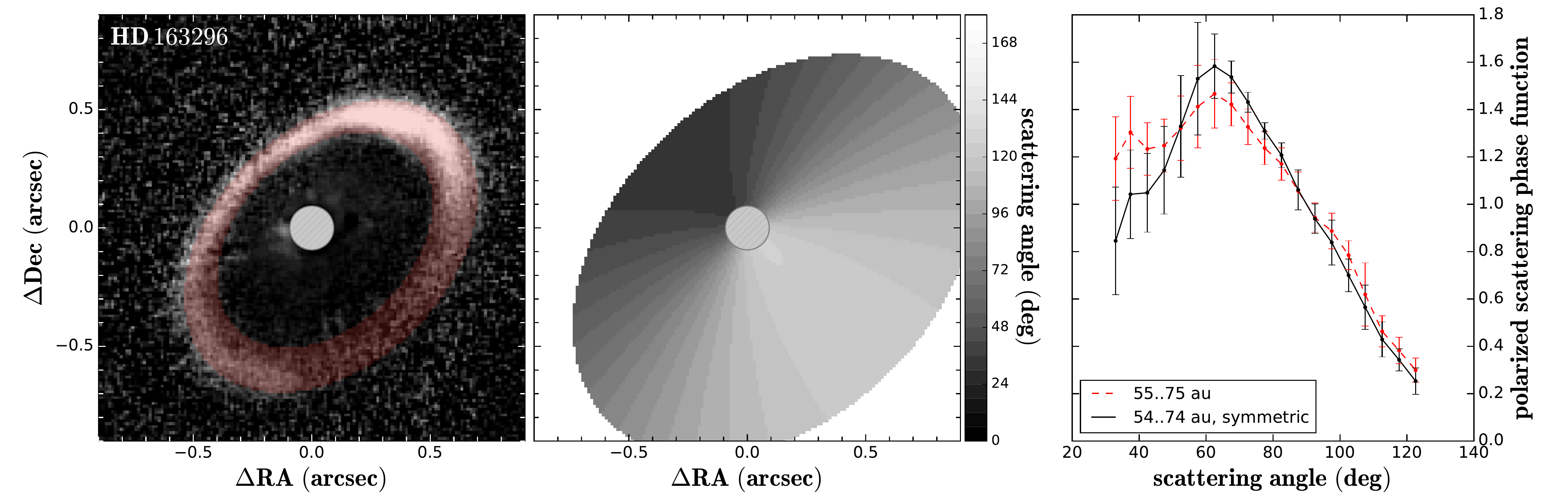} 
\caption{Extracted polarized scattered phase functions and extraction regions for the H-band data of the HD\,163296 system. Panels are analog to figure~\ref{fig:rxj1615}.
} 
\label{fig:hd163296}
\end{figure*}

\begin{figure*}
\center
\includegraphics[width=0.99\textwidth]{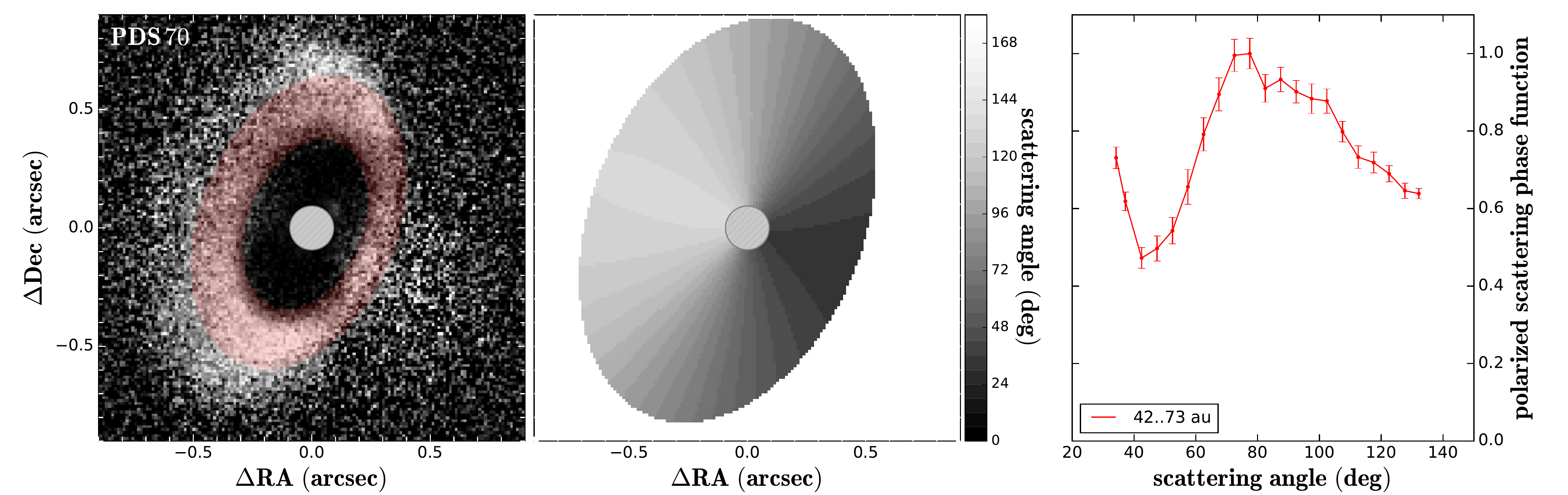} 
\caption{Extracted polarized scattered phase functions and extraction regions for the H-band data of the PDS\,70 system. Panels are analog to figure~\ref{fig:rxj1615}.
} 
\label{fig:pds70}
\end{figure*}

\begin{figure*}
\center
\includegraphics[width=0.79\textwidth]{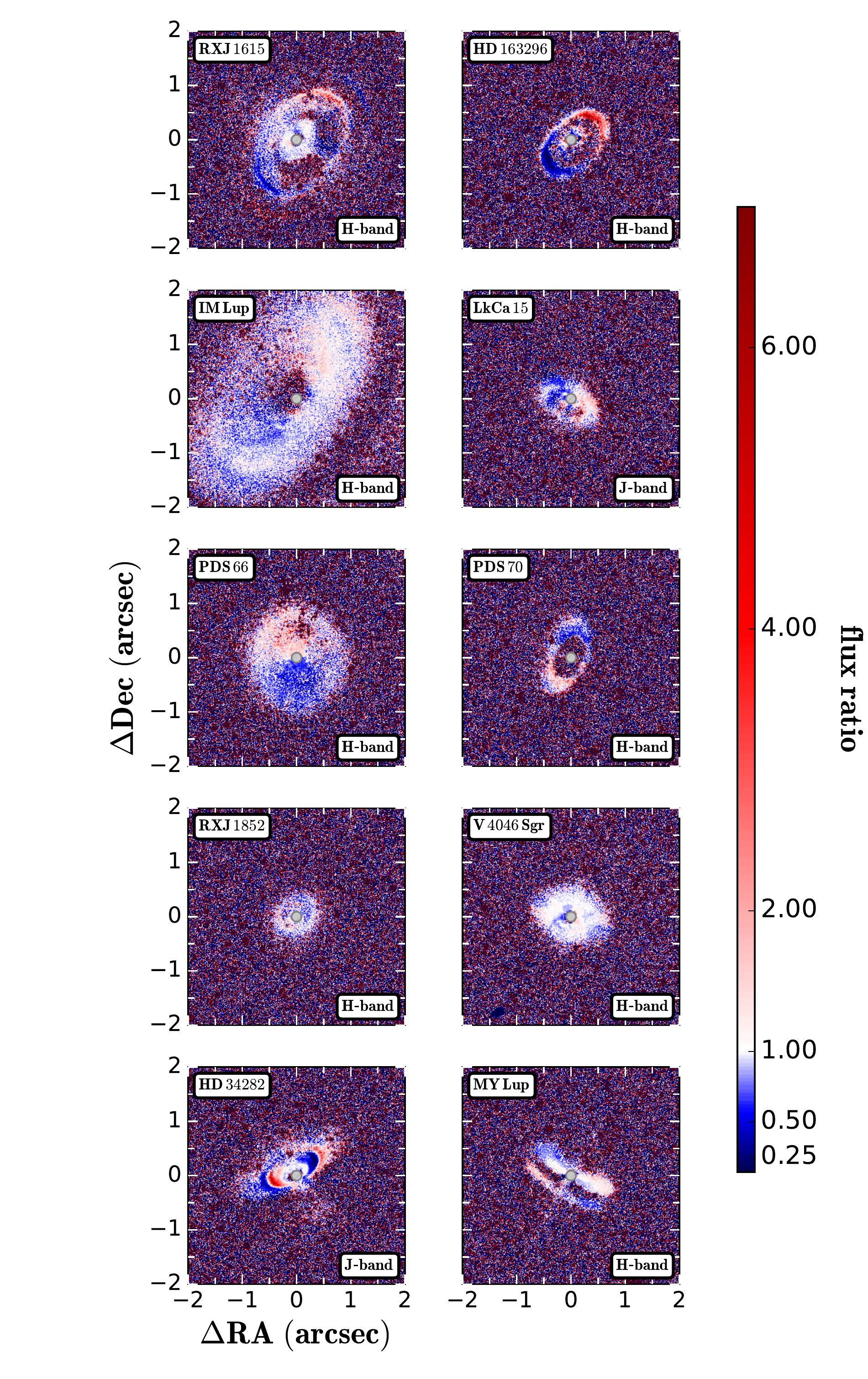} 
\caption{Axis symmetry of all disks in our sample relative to the disk minor axis. H-band data is shown when available otherwise J-band data. The disk images were flipped around the minor axis and then the original images was divided by the flipped image. Thus flux ratios$>$1 indicate the factor by which the disk region is brighter than the corresponding axis-symmetric region.   
} 
\label{fig:shadow gallery}
\end{figure*}

\section{$T$-matrix calculations} \label{sec:tmatrix}

We considered four different types of dust aggregation: Ballistic Cluster-Cluster Aggregation (BCCA); Ballistic Particle-Cluster Aggregation (BPCA), and two modified versions of BPCA, known as BAM1 and BAM2 \citep{Shen08}. BCCA has a fractal dimension of 1.9 and therefore has a highly open structure, whereas the other three have a fractal dimension close to three. The main difference between BPCA, BAM1, and BAM2 are their porosities; the lowest for BAM2. 
We assume a spherical and single-sized monomer for the computational convenience. Each monomer has a dust composition with a mixture of water ice \citep{Warren08}, pyroxene silicate (Mg$_{0.7}$Fe$_{0.3}$SiO$_3$) \citep{Dor95}, amorphous carbon (\citealt{Zubko1996}), and troilite \citep{Henning96} with the mass abundance ratios similar to the DSHARP model \citep{Birnstiel18}. We calculated the effective refractive index ($m$) by using the Bruggeman mixing rule and found $m=1.92+0.404i$ at a wavelength of $1.63~\mu$m.

Given dust geometry and composition, we calculate the scattering matrix elements of dust aggregates using by Multiple Sphere $T$-Matrix Method \citep{Mackowski11}. In the simulations, we assume that aggregates are randomly orientated, i.e.., ignoring grain alignment, and their optical properties were averaged over all possible orientation with equal probability by using the analytical orientation averaging technique of the $T$-matrix method. The results were also averaged over four realizations of each aggregate model.

Once we obtain the optical properties for each aggregate, we then averaged the optical properties by considering aggregate-size distribution:
\begin{equation}
n(a)da\propto a^{-3.5}da~(\amin\le a \le \amax),
\end{equation}
where $a$ is the volume-equivalent radius of an aggregate, defined by $a=\amon N^{1/3}$; $\amon$ being the radius of the monomer and $N$ being the number of monomers. The minimum aggregate radius is fixed to $\amin=2\amon$, and the maximum aggregate radius $\amax$ is a parameter. We consider two different monomer radii: $\amon=0.1~\mu$m and $0.4~\mu$m. The largest aggregates we investigated have $\amax=1.6~\mu$m. Since the volume-equivalent radius does not necessarily represents the apparent size of an aggregate, we introduce the characteristic radius $a_\mathrm{c}$ \citep{Mukai92}, which better describes the apparent size. We measure the porosity by $\mathcal{P}=1-(a/a_c)^3$. The characteristic radius and porosity of the maximum aggregate in the size distribution will be denoted by $a_\mathrm{c,max}$ and $\mathcal{P}_\mathrm{max}$, respectively.

\end{appendix}

\end{document}